\documentclass[prd,12pt]{article}

\usepackage{amsmath,amssymb,graphicx,multirow,xspace,slashed,array,booktabs}
\usepackage[colorlinks=true,urlcolor=blue,anchorcolor=blue,citecolor=blue,filecolor=blue,linkcolor=blue,menucolor=blue]{hyperref}
\usepackage[compress,numbers]{natbib}
\usepackage[labelformat=simple]{subcaption}

\usepackage{booktabs}
\usepackage{blindtext, rotating}
\usepackage{afterpage}
\usepackage{enumitem}
\usepackage{marvosym}
\usepackage{authblk}
\usepackage{verbatim}
\usepackage{multirow}
\usepackage{soul} 
\usepackage[normalem]{ulem}
\usepackage{pifont}
\usepackage{booktabs}
\usepackage{bbm}
\usepackage{tikz,xcolor}
\usepackage{caption}
\usepackage{subcaption}
\usepackage{amsmath}
\usepackage{slashed}
\usepackage{comment}
\usepackage{cancel}
\usepackage{multirow}
\usepackage{tikz-feynman}
\usepackage{orcidlink}
\tikzfeynmanset{compat=1.1.1}
\usepackage{footmisc}

\usepackage{floatrow}
\newfloatcommand{capbtabbox}{table}[][\FBwidth]

\usepackage[font=footnotesize,labelfont=bf]{caption}

\usepackage{lineno}

\allowdisplaybreaks

\usepackage[font=footnotesize,labelfont=bf]{caption}

\newcommand{\bea}{\begin{array}}
\newcommand{\eea}{\end{array}}
\newcommand{\beq}{\begin{eqnarray}}
\newcommand{\eeq}{\end{eqnarray}}

\numberwithin{equation}{section}

\newcommand{\slg}[1]{#1\hspace{-0.45em}/}

\newcommand{\DarkFac}{ \mathbb{F}_D}

\begin{document}
\baselineskip 0.6cm

\begin{titlepage}

\vspace*{-1.5cm}

\thispagestyle{empty}

\begin{flushright}
CTPU-PTC-26-14
\end{flushright}

\begin{center}
\vskip 1cm

{\large \bf Neutron Portal and Dark Matter-Baryon Coincidence:\\[0.3ex]
from UV Completion to Phenomenology}
\vskip 1cm

{\large Sudhakantha Girmohanta \orcidlink{0000-0003-4209-1118} $^{1}$, Yuichiro Nakai \orcidlink{0000-0002-3547-3145} $^{2,3}$, \\[1ex]
Yoshihiro Shigekami \orcidlink{0000-0001-7354-5018} $^{4}$, and Zhihao Zhang \orcidlink{0009-0003-8073-680X} $^{2,3}$}
\vskip 0.5cm
{\it
$^1$Particle Theory and Cosmology Group, Center for Theoretical Physics of the Universe, Institute for Basic Science (IBS), Daejeon, 34126, Korea \\
$^2$Tsung-Dao Lee Institute, Shanghai Jiao Tong University, \\
No.~1 Lisuo Road, Pudong New Area, Shanghai, 201210, China \\
$^3$School of Physics and Astronomy, Shanghai Jiao Tong University, \\
800 Dongchuan Road, Shanghai, 200240, China \\
$^4$School of Physics, Henan Normal University, Xinxiang, 453007, Henan, China}
\vskip 1.0cm
\end{center}

\begin{abstract}

We present a dynamical solution to the dark matter-baryon coincidence problem based on the neutron portal operator connecting the visible and dark sector asymmetries. 
This framework is motivated by the possibility that a strongly supercooled dark confinement phase transition accounts for the nano-Hz stochastic gravitational wave signal observed by pulsar timing arrays, while also generating the dark matter and baryon asymmetry in the Universe. 
We show that the GeV-scale mass of asymmetric dark matter can be naturally correlated with the (multi-)TeV scale cut-off for the neutron portal through its ultraviolet completion. 
The dark sector is governed by an approximate fixed point and confines once the heavy portal states are integrated out, dynamically generating a scale of $\mathcal{O} ({\rm GeV})$. 
We analyze both tree and loop-level ultraviolet completions and demonstrate how the resulting confinement scale is linked to the effective neutron portal scale. 
We also discuss cosmological constraints and experimental prospects in beam dump searches and colliders for probing the neutron portal. 

\end{abstract}

\flushbottom

\end{titlepage}



\section{Introduction}
\label{sec:introduction}

The asymmetric dark matter (ADM) paradigm~\cite{Nussinov:1985xr,Barr:1990ca,Barr:1991qn,Kaplan:1991ah,Gudnason:2006ug,Gudnason:2006yj,Davoudiasl:2012uw,Petraki:2013wwa,Shelton:2010ta,Zurek:2013wia} is motivated by the ``dark matter-baryon coincidence puzzle'', which is the observation that the present dark matter (DM) and baryon abundances are not orders of magnitude apart, despite possible distinct cosmological origins, namely
\begin{equation}
\Omega_{\rm DM} = 0.256(7) \, , \quad \Omega_{\rm b} = 0.049(3) \, ,
\label{Eq:Omegas}
\end{equation}
where $\Omega_{{\rm DM}, {\rm b}}$ correspond to the energy density fractions of DM and visible matter, respectively.\footnote{The abundance for fluid $i$ having energy density $\rho_i$ is defined as $\Omega_i \equiv \rho_i / \rho_{\rm crit}$, with $\rho_{\rm crit} = 3 H_0^2 / (8 \pi G)$, where $H_0$ is the current Hubble constant and $G$ is the Newton gravitational constant~\cite{ParticleDataGroup:2024cfk}.} 
In this framework, a primordial asymmetry between the number densities of DM and anti-DM is related to the visible baryon asymmetry through some portal. 
As the asymmetries share the same origin, it is expected that 
\begin{equation}
n_{\rm DM} - n_{\overline{\rm DM}} \simeq n_{\rm b} - n_{\overline {\rm b}} \, ,
\label{Eq:asymm}
\end{equation}
where $n_i$ represents the number density in species $i$. 
The symmetric component of the DM annihilates away in the early Universe, latest before the structure formation, leaving the asymmetric part intact, which constitutes the present DM in the Universe. 
Therefore, explaining the DM abundance requires
\begin{equation}
\frac{\Omega_{\rm DM}}{\Omega_{\rm b}} = \frac{m_{\rm DM} (n_{\rm DM} - n_{\overline{\rm DM}})}{m_p (n_{\rm b} - n_{\overline {\rm b}})} \simeq \frac{m_{\rm DM}}{m_p} \simeq 5.4 \, ,
\label{Eq:ratio}
\end{equation}
where $m_{\rm DM}$ is the DM mass, and $m_p \simeq 0.938 \, {\rm GeV}$ is the mass of the proton. 
Therefore, to explain the coincidence satisfactorily, one has to also explain why the DM mass lies in the GeV range. 

Another hint for a GeV-scale dark sector (DS) emerges from the possibility of explaining the nano-Hz stochastic gravitational waves (GWs) observed by the pulsar timing array (PTA) collaboration~\cite{NANOGrav:2023gor,NANOGrav:2023hvm,EPTA:2023fyk,Reardon:2023gzh,Xu:2023wog} from a first-order phase transition (PT) in the DS~\cite{Nakai:2020oit,Fujikura:2023lkn,Madge:2023dxc,Megias:2023kiy,Salvio:2023ynn,Salvio:2023blb,Gouttenoire:2023bqy,Addazi:2023jvg,Li:2023bxy,Ghosh:2023aum,Jiang:2023qbm,Wang:2023bbc,Li:2025nja,Fujikura:2025iam,Chatrchyan:2025wop,Musumeci:2026jum}. 
The nano-Hz peak frequency of the observed GW, when properly accounting for the expansion history of the Universe, corresponds to a PT with reheating temperature in the GeV scale. 
Concretely, we consider a DS governed by a nearly conformal dynamics in the ultraviolet (UV) that undergoes a first-order confinement–deconfinement PT triggered by the mass gap generated by a confining dark QCD dynamics~\cite{Fujikura:2023lkn}. 
The nucleation and subsequent collisions of true-vacuum bubbles generate a stochastic gravitational-wave background in the nano-Hz frequency range, which can account for the PTA signal. 
Moreover, such a PT can provide a better fit to the observed GW spectral shape than the baseline supermassive black hole inspiral scenario~\cite{Ellis:2023oxs}. 
However, to fit the PTA signal, the phase transition must be strongly supercooled, leading to substantial entropy production during reheating~\cite{Madge:2023dxc,Ellis:2020nnr}. 
This entropy release dilutes any baryon asymmetry and dark matter abundance existing before this GeV-scale PT. 
Consequently, creating the baryon asymmetry and DM utilizing the PT itself is appealing, which naturally links this scenario to the ADM framework~\cite{Fujikura:2024jto,Girmohanta:2025wcq}. 
Here, the lightest dark baryon plays the role of the DM, whose mass is set by the confinement scale of the dark QCD, denoted by $\Lambda_{\rm dQCD}$. 
Thus, both the coincidence puzzle and the PTA signal can be simultaneously addressed provided that the confinement scale satisfies $\Lambda_{\rm dQCD} \sim \mathcal{O} ({\rm GeV})$. 
As an additional virtue, the composite DM is naturally self-interacting through the exchange of dark pions, and a confinement scale $\Lambda_{\rm dQCD} \sim \mathcal{O} ({\rm GeV})$ may yield the desired self-interaction cross-section to explain the observed diversity of the inner slopes of galactic rotation curves~\cite{Cline:2013zca,Tulin:2017ara,Roberts:2024uyw,Zeng:2024xty,Chung:2025wle}. 
Explaining the emergence of the GeV scale in the DS in the framework of ADM is the goal of the present work. 

A necessary ingredient for the ADM scenario to work is the existence of a portal that shares the asymmetry between the dark and visible sectors. 
The asymmetry might be created in the DS through some processes that violate a global dark number $U(1)_D$, e.g., through anomaly~\cite{Fujikura:2024jto}, or the decay of a heavy Majorana particle~\cite{Girmohanta:2025wcq}. 
This dark asymmetry is then reprocessed into the visible sector via a portal operator that violates both $U(1)_D$ and Standard Model (SM) baryon ($B$) or (total) lepton number ($L$). 
If the relevant temperature of the process is above the electroweak scale, then numerous effective operators/electroweak sphalerons can be utilized to reprocess this asymmetry. 
On the other hand, if the relevant temperature is below $\simeq 130 \, {\rm GeV}$, as is the case for the PT explaining the PTA, then the sphalerons are frozen out, and a portal has to be used which directly violates baryon number $B$. 
The lowest dimensional effective operator relevant for this purpose is known as the \textit{neutron portal} operator\footnote{If quark fields from higher generations are included in the portal operator, additional decay channels are induced, such as heavy baryon $\to$ meson$+ \bar{\chi}$ and heavy meson $\to$ baryon$+ \bar{\chi}$~\cite{Heeck:2020nbq}. 
As a concrete example, constraints from BABAR reported in Ref.~\cite{BaBar:2023dtq} impose a lower bound on the cutoff scale of the $u d b \chi$ operator of order $\Lambda_{u d b} \gtrsim 6 \, {\rm TeV}$ for $m_{\chi} \sim \mathcal{O} ({\rm GeV})$. 
We restrict ourselves to the first-generation quarks for the current analysis.}
\begin{equation}
{\cal O}_{n \chi} = \frac{1}{\Lambda_n^2} \left( \overline{\chi^c} d_R^c \right) \left( \overline{u_R} d_R^c \right) \, ,
\label{Eq:portal}
\end{equation}
where $\chi$ is a DS fermion, and carries a non-zero $U(1)_D$, $\Lambda_n$ is the effective cut-off scale, $c$ denotes charge conjugation, and the color indices are suppressed. 
Asymmetry in $\chi$ number density is created during the PT, which is communicated to the DM and visible baryons. 
We assume $\chi$ to be heavier than a neutron to respect nucleon decay constraints. 
To ensure that $\chi$ decays before the onset of Big Bang Nucleosynthesis (BBN) requires $\Lambda_n \lesssim \mathcal{O} (100) \, {\rm TeV}$, while if one demands that this operator remains in equilibrium at the GeV temperature, then $\Lambda_n \lesssim 15 \, {\rm TeV}$. 
The latter constraint is not necessary for the asymmetry sharing, as $\chi$ decay can carry the asymmetry to the visible baryons.\footnote{The phase transition may induce plasma inhomogeneities which could impact BBN~\cite{Applegate:1987hm,Bagherian:2025puf}. 
Since the asymmetry is first generated in the DS and later transferred to visible baryons, a dedicated study including DS diffusion is required, which we leave for a future exploration.} 

In our present work, we ponder if the existence of this (multi-)TeV $\Lambda_n$ is connected to the emergence of the GeV scale in the DS as follows: to UV complete the effective operator in Eq.~\eqref{Eq:portal}, new color-charged and dark QCD charged particles have to be introduced, depending on whether the UV completion is realized at the tree or loop-level. 
If the theory is governed by an approximate infrared (IR) fixed point, once these new states obtain masses and are integrated out, the dark QCD flows away from the fixed point and confines. 
We analyze when integrating out these new states results into $\Lambda_{\rm dQCD} = \mathcal{O} (\rm GeV)$, and correlate it to the corresponding $\Lambda_n$. 
We also outline how the masses of these new particles may originate in the context of a solution for the $\mu$ problem in the supersymmetric extension of the SM. 

Previous studies addressing the coincidence problem invoked exact or partially broken mirror symmetry~\cite{Hodges:1993yb,Foot:2003jt,An:2009vq,Farina:2015uea,Lonsdale:2018xwd,Ibe:2018juk,Bodas:2024idn}, dark unification with QCD~\cite{Ibe:2018tex,Ibe:2019ena,Murgui:2021eqf,Chung:2024nnj}, IR fixed-point dynamics~\cite{Bai:2013xga,Newstead:2014jva,Ritter:2022opo,Ritter:2024sqv}, or some other related ideas~\cite{Brzeminski:2023wza,Borah:2024wos,Banerjee:2024xhn,Chung:2024ezq,Cox:2025wxk}. 
While our proposal also falls within the class of fixed-point dynamics, it features several key differences. 
We introduce no ad-hoc QCD or dark QCD charged matter beyond what is required for a UV completion of the neutron portal operator, contrary to the arbitrary new particle contents in the previous studies. 
The QCD coupling is allowed to run slowly in the far UV (so that our scenario is consistent with the ordinary grand unification framework), whereas the dark QCD coupling exhibits fixed-point behavior, leading to a correlation between $\Lambda_{\rm dQCD}$ and $\Lambda_n$. 
Proceeding further, we study the reach of current and future beam-dump experiments in probing the neutron portal, compare it with the jet plus missing energy searches in colliders, and analyze BBN and Cosmic Microwave Background (CMB) constraints for $m_{\chi} \gtrsim {\rm GeV}$. 

The rest of the paper is organized as follows. 
Section~\ref{Sec:models} introduces the tree and loop-level UV completion of the neutron portal and analysis of the fixed-point dynamics. 
In section~\ref{Sec:darkpion}, we discuss the phenomenology of a dark baryon DM. 
Section~\ref{Sec:phenomenology} provides a phenomenological analysis of the neutron portal operator, including beam dump, collider searches, and constraints from BBN and CMB. 
Section~\ref{Sec:conclusion} contains our conclusions. 
Some details are summarized in appendices.

\section{UV completion of the neutron portal}
\label{Sec:models}

In order to explain the coincidence problem of $\Omega_{\rm DM} \simeq 5.4 \Omega_{\rm b}$ in the context of the ADM model, one has to address why the DM mass is $m_{\rm DM} \simeq 5 \, {\rm GeV}$. 
In this section, we would like to illustrate how the GeV scale may originate in connection with the UV completion of the neutron portal. 
For this purpose, we present two cases to reproduce the neutron portal operator by introducing new particles. 
Then, we discuss the connection between the neutron portal and a DS, like a dark QCD.

\subsection{Tree-level neutron portal}
\label{sec:UV-tree}

For reproducing the neutron portal operator in Eq.~\eqref{Eq:portal}, new SM colored particles should couple to $u, d$ quarks and also to $\chi$.\footnote{We consider $\chi$ to be purely Dirac in nature, although it is totally singlet under the SM gauge symmetries. 
If $\chi$ has an effective Majorana mass term, the neutron portal operator will give rise to $n$--$\bar{n}$ oscillations and can therefore be constrained by searches for them~\cite{Girmohanta:2025wcq,McKeen:2015cuz}.} 
The simplest way to realize this situation is to introduce one colored-scalar $\Phi$ in the model, with following Lagrangian:
\begin{align}
- \mathcal{L} \supset y_{q \Phi} \epsilon^{\alpha \beta \gamma} \overline{u_R}_{\alpha} d_{R \, \beta}^c \Phi^*_{\gamma} + y_{\chi \Phi} \overline{\chi^c} d_{R \, \alpha}^c \Phi^{\alpha} + {\rm h.c.} \, ,
\label{eq:Lag-tree}
\end{align}
where $y_{q \Phi}, y_{\chi \Phi}$ are assumed to be real and positive, $\alpha, \beta, \gamma$ are $SU(3)_C$ indices, $SU(3)_C$ charge of $\Phi$ is $\mathbf{3}$ and its $U(1)_{\rm em}$ charge is $- \frac{1}{3}$. 
Note that in this work, $\Phi$ as well as new particles introduced below are $SU(2)_L$ singlets. 
The charge conjugation is defined by
\begin{align}
d^c_R = C \overline{d_R}^T ~~ \text{with } C \equiv i \gamma^2 \gamma^0 \, .
\label{eq:defCC}
\end{align}
In this model, the neutron portal operator can be obtained from the ``tree-level" process, mediated by $\Phi$. 
It is notable that we can consider the other types of Lagrangian for the tree-level process: (i) the case with $u_R \leftrightarrow d_R$ in the first term and others keeping unchanged, (ii) the case with $\overline{u_R} \to \overline{d_R}$ in the first term, $d_R^c \to u_R^c$ in the last term, and $- \frac{1}{3} \to + \frac{2}{3}$ for $U(1)_{\rm em}$ charge of $\Phi$. 
The case (i) is equivalent to Eq.~\eqref{eq:Lag-tree}, since
\begin{align}
\epsilon^{\alpha \beta \gamma} \overline{d_R}_{\alpha} u_{R \, \beta}^c \Phi_{\gamma}^* = \epsilon^{\alpha \beta \gamma} \overline{u_R}_{\beta} d_{R \, \alpha}^c \Phi_{\gamma}^* = - \epsilon^{\alpha \beta \gamma} \overline{u_R}_{\alpha} d_{R \, \beta}^c \Phi_{\gamma}^*
\label{eq:CCandSU(3)}
\end{align}
with use of definition in Eq.~\eqref{eq:defCC} and anti-symmetric property of $\epsilon^{\alpha \beta \gamma}$, and hence, only the sign of $y_{q \Phi}$ is changed. 
However, for the case (ii), the first term is vanished by applying Eq.~\eqref{eq:CCandSU(3)}: $\epsilon^{\alpha \beta \gamma} \overline{d_R}_{\alpha} d_{R \, \beta}^c \Phi_{\gamma}^* = - \epsilon^{\alpha \beta \gamma} \overline{d_R}_{\alpha} d_{R \, \beta}^c \Phi_{\gamma}^*$. 
As a result, it is sufficient to consider Eq.~\eqref{eq:Lag-tree} for the neutron operator induced by the tree-level process. 
Using two Yukawa couplings in Eq.~\eqref{eq:Lag-tree}, we can estimate $\Lambda_n$ in Eq.~\eqref{Eq:portal} as
\begin{align}
\left. \Lambda_n \right|_{\rm tree} \simeq \frac{m_{\Phi}}{\sqrt{y_{q \Phi} y_{\chi \Phi}}} \, ,
\label{eq:Lamn-tree}
\end{align}
where we assume that the four momentum of $\Phi$ in the propagator is much smaller than the mass of $\Phi$, denoted as $m_{\Phi}$. 

Although this tree-level process is the simplest case and leads to a minimal model, $\Phi$ cannot have any dark charges, for e.g., under the dark QCD. 
Hence, it is not apparent how the GeV-scale in the DS appears. 
Nevertheless, $\left. \Lambda_n \right|_{\rm tree}$ can be related to some information about the DS, through the mass of $\Phi$. 
For example, $\Phi$ can couple to some dark scalar, $\phi_D$ through $\lambda_{\Phi \phi_D} |\Phi|^2 |\phi_D|^2$ in the scalar potential, and once $\phi_D$ acquires non-zero vacuum expectation value (VEV) denoted as $v_D$, the mass of $\Phi$ has a contribution from this term, $m_{\Phi}^2 \supset \lambda_{\Phi \phi_D} v_D^2$. 
If $m_{\Phi}^2$ is dominated by this $\lambda_{\Phi \phi_D} v_D^2$ term, $\left. \Lambda_n \right|_{\rm tree}$ is determined by $v_D$ with $\mathcal{O} (1)$ couplings of $y_{q \Phi}, y_{\chi \Phi}, \lambda_{\Phi \phi_D}$, and hence, $2 \, {\rm TeV} \lesssim \Lambda_n \lesssim 15 \, {\rm TeV}$ can be simply reproduced when $2 \, {\rm TeV} \lesssim v_D \lesssim 15 \, {\rm TeV}$. 
$\phi_D$ on the other hand, can couple to DS fermions that may obtain mass at the scale of $\sim v_D$ and therefore, $\left. \Lambda_n \right|_{\rm tree}$ can be indirectly related to the appearance of the GeV-scale in the dark sector, once these TeV-scale particles are integrated out and the dark QCD confines around the GeV scale. 
At the same time, by model building, it can be ensured that some light dark quarks do not get TeV-scale mass that forms dark baryon DM after dark QCD confines, such that DM mass is primarily dictated by $\Lambda_{\rm dQCD}$. 
For example, one can consider a global or gauged dark symmetry under which the light dark quark is chiral, and $\phi_D$ is not charged.

\subsection{Loop-level neutron portal}
\label{sec:UV-1loop}

\begin{figure}[t]
\begin{center}
\begin{tikzpicture}
\begin{feynman}[large]
\vertex (a1) {\(d\)};
\vertex [above=3.4cm of a1] (d1) {\(u\)};
\vertex [above=1.7cm of a1] (ma1);
\vertex[blob,shape=ellipse,minimum height=0.8cm,minimum width=0.8cm] [right=0.2cm of ma1] (vl1) {};
\vertex [right=3.0cm of a1] (b1) {\(\chi\)};
\vertex [above=3.4cm of b1] (c1) {\(d\)};
\vertex [above=1.7cm of b1] (mb1);
\vertex [left=0.6cm of mb1] (vr1);
\vertex [right=1.5cm of a1] (mc1);
\vertex [below=0.3cm of mc1] (l1) {(a)};
\diagram [medium] {
(a1) -- (vl1) -- (d1),
(vl1) -- [scalar, edge label=\(\Phi\)] (vr1),
(c1) -- (vr1) -- (b1),
};
\vertex [right=1.1cm of b1] (a2) {\(d\)};
\vertex [above=3.4cm of a2] (d2) {\(u\)};
\vertex [above=1.7cm of a2] (ma2);
\vertex [right=0.6cm of ma2] (vl2);
\vertex [right=3.0cm of a2] (b2) {\(\chi\)};
\vertex [above=3.4cm of b2] (c2) {\(d\)};
\vertex [above=1.7cm of b2] (mb2);
\vertex [right=1.5cm of a2] (mc2);
\vertex [below=0.3cm of mc2] (l2) {(b)};
\vertex[blob,shape=ellipse,minimum height=0.8cm,minimum width=0.8cm] [left=0.2cm of mb2] (vr2) {};
\diagram [medium] {
(a2) -- (vl2) -- (d2),
(vl2) -- [scalar, edge label=\(\Phi'\)] (vr2),
(c2) -- (vr2) -- (b2),
};
\vertex [right=1.1cm of b2] (a3) {\(d\)};
\vertex [above=3.4cm of a3] (d3) {\(u\)};
\vertex [above=1.7cm of a3] (ma3);
\vertex [right=0.6cm of ma3] (vl3);
\vertex[blob,shape=ellipse,minimum height=0.6cm,minimum width=0.6cm] [right=0.6cm of vl3] (vm3) {};
\vertex [right=3.0cm of a3] (b3) {\(\chi\)};
\vertex [above=3.4cm of b3] (c3) {\(d\)};
\vertex [above=1.7cm of b3] (mb3);
\vertex [left=0.6cm of mb3] (vr3);
\vertex [right=1.5cm of a3] (mc3);
\vertex [below=0.3cm of mc3] (l3) {(c)};
\diagram [medium] {
(a3) -- (vl3) -- (d3),
(vl3) -- [scalar, edge label=\(\Phi'\)] (vm3) -- [scalar, edge label=\(\Phi\)] (vr3),
(c3) -- (vr3) -- (b3),
};
\vertex [right=1.1cm of b3] (a4) {\(d\)};
\vertex [above=3.4cm of a4] (d4) {\(u\)};
\vertex [above=1.7cm of a4] (ma4);
\vertex[blob,shape=ellipse,minimum height=1.0cm,minimum width=1.0cm] [right=1.0cm of ma4] (vm4) {};
\vertex [right=3.0cm of a4] (b4) {\(\chi\)};
\vertex [above=3.4cm of b4] (c4) {\(d\)};
\vertex [right=1.5cm of a4] (mc4);
\vertex [below=0.3cm of mc4] (l4) {(d)};
\diagram [medium] {
(a4) -- (vm4) -- (d4),
(b4) -- (vm4) -- (c4),
};
\end{feynman}
\end{tikzpicture}
\end{center}
\vspace{-0.7cm}
\caption{Possible loop diagrams which generate the neutron portal operator. 
Each blob has a loop of dark charged particles, while $\Phi$ and $\Phi'$ are not charged under any dark symmetries. }
\label{fig:diags-LoopAll}
\end{figure}
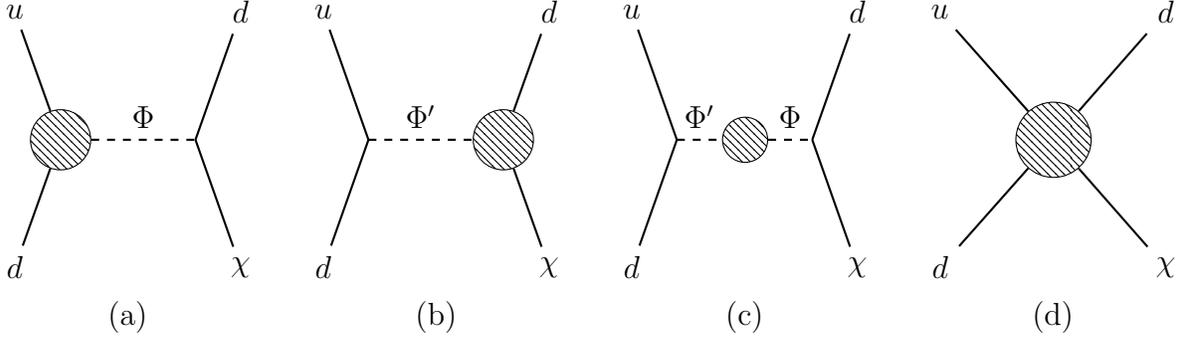

For the other possibility to reproduce the neutron operator, we can consider loop diagrams of dark charged particles. 
Hereafter, we call this ``loop-level" process.\footnote{One can also consider the operator of the form $\bar{\chi} d_R Q_L Q_L$, which leads to a bit different loop-level UV completion. 
However, it gives similar phenomenology, and is also less relevant for the supersymmetric case discussed later. 
Therefore, we focus on the operator structure in Eq.~\eqref{Eq:portal}.} 
We have four possible diagrams for the loop-level process, shown in Fig.~\ref{fig:diags-LoopAll}. 
The blob in each diagram indicates the loop of dark charged particles. 
Diagrams (a), (b) and (c) require new particle $\Phi$ and/or $\Phi'$ which are singlet under the dark symmetries, as introduced in the case of the tree-level process, while the diagram (d) does not need this kind of particle. 
Before discussing the detail of this loop, we can consider the $SU(3)_C$ representations of $\Phi$ and $\Phi'$. 

For diagram (a), $\Phi$ should be fundamental representation $\mathbf{3}$ due to $\Phi$-$d$-$\chi$ coupling, which is the same case to the tree-level process. 
Therefore, when we consider the loop diagram (a), we always have the tree-level process, because there is no reason to ignore it by any symmetries. 
As a result, the contribution from loop diagram (a) is sub-dominant. 
The situation is the same for diagram (c). 

The diagram (b) is a different case from diagrams (a) and (c), in the sense of allowed representation for $\Phi'$. 
Since $\Phi'$ couples to $u$ and $d$, we can choose $\Phi'$ to be $\mathbf{\bar{6}}$ instead of $\mathbf{3}$, because of the $SU(3)_C$ tensor product rule: $\mathbf{3} \otimes \mathbf{3} = \mathbf{\bar{3}} \oplus \mathbf{6}$. 
In this case, the tree-level process is obviously forbidden, and the diagram (b) will be dominant loop contribution to the neutron portal operator. 
However, $\mathbf{\bar{6}}$ has symmetric for its $SU(3)_C$ indices, $\mathbf{\bar{6}}^{\alpha \beta} = \mathbf{\bar{6}}^{\beta \alpha}$, and these indices should be contracted with two SM quarks, $u$ and $d$. 
This conflicts with the final form of the neutron portal operator: to construct the neutron state, all three SM quarks should be anti-symmetric combination. 
We checked that in the actual calculation of the loop part, the totally anti-symmetric tensor of $SU(3)_C$, $\epsilon^{\alpha \beta \gamma}$, appears, whose two indices are contracted with those of $\Phi'$. 
This results in vanishing amplitude, namely, $\epsilon^{\alpha \beta \gamma} \Phi^{\prime}_{\alpha \beta} = 0$. 

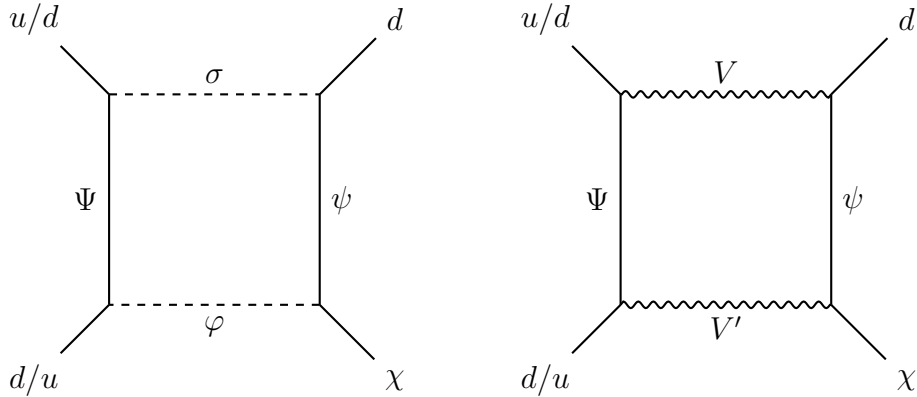
\begin{figure}[t]
\begin{center}
\begin{tikzpicture}
\begin{feynman}[large]
\vertex (a5) {\(d/u\)};
\vertex [above=4.8cm of a5] (d5) {\(u/d\)};
\vertex [above=1.0cm of a5] (me1);
\vertex [right=1.0cm of me1] (me2);
\vertex [below=1.0cm of d5] (me3);
\vertex [right=1.0cm of me3] (me4);
\vertex [right=4.8cm of a5] (b5) {\(\chi\)};
\vertex [above=4.8cm of b5] (c5) {\(d\)};
\vertex [above=1.0cm of b5] (me5);
\vertex [left=1.0cm of me5] (me6);
\vertex [below=1.0cm of c5] (me7);
\vertex [left=1.0cm of me7] (me8);
\vertex [right=2.4cm of a5] (me9);
\diagram [medium] {
(a5) -- (me2) -- [edge label=\(\Psi\)] (me4) -- (d5),
(c5) -- (me8) -- [edge label=\(\psi\)] (me6) -- (b5),
(me6) -- [scalar, edge label=\(\varphi\)] (me2),
(me4) -- [scalar, edge label=\(\sigma\)] (me8),
};
\vertex [right=2.0cm of b5] (a6) {\(d/u\)};
\vertex [above=4.8cm of a6] (d6) {\(u/d\)};
\vertex [above=1.0cm of a6] (mf1);
\vertex [right=1.0cm of mf1] (mf2);
\vertex [below=1.0cm of d6] (mf3);
\vertex [right=1.0cm of mf3] (mf4);
\vertex [right=4.8cm of a6] (b6) {\(\chi\)};
\vertex [above=4.8cm of b6] (c6) {\(d\)};
\vertex [above=1.0cm of b6] (mf5);
\vertex [left=1.0cm of mf5] (mf6);
\vertex [below=1.0cm of c6] (mf7);
\vertex [left=1.0cm of mf7] (mf8);
\diagram [medium] {
(a6) -- (mf2) -- [edge label=\(\Psi\)] (mf4) -- (d6),
(c6) -- (mf8) -- [edge label=\(\psi\)] (mf6) -- (b6),
(mf6) -- [boson, edge label=\(V'\)] (mf2),
(mf4) -- [boson, edge label=\(V\)] (mf8),
};
\end{feynman}
\end{tikzpicture}
\end{center}
\vspace{-0.5cm}
\caption{Loop diagrams which generate the neutron portal operator. 
Here, we omit diagrams with the fermion line of $d$-$\Psi$-$d$, because this diagram has vanishing amplitude, as we explained above (see Eq.~\eqref{eq:CCandSU(3)} and discussion around that). }
\label{fig:diags-Loop}
\end{figure}

As a result, diagrams (a), (b) and (c) are less important to discuss the neutron portal operator generated by the loop-level process: their contributions cannot be dominant, since they either admit a tree-level UV completion of the neutron portal operator or vanish due to symmetry properties.\footnote{For the other possibility, one can consider the diagram with two blobs, namely, $u$-$d$-$\tilde{\Phi}$ and $d$-$\chi$-$\tilde{\Phi}$ vertices are induced by dark particle loops. 
This will be dominant contribution to the neutron portal operator, because the new particles $\tilde{\Phi}$ connecting two blobs can have dark charges, which lead the tree-level process to be vanishing. 
However, this diagram appears at the two-loop or higher loop order, and a resultant neutron portal scale $\Lambda_n$ tends to be higher than $\mathcal{O} (1) \, {\rm TeV}$ due to additional loop factor, as we will see later. 
Therefore, we ignore this diagram in this work.} 
Then, we focus on the last diagram (d). 
Since all external fields are fermions, this kind of diagrams can be obtained by a box-type diagram. 
Therefore, one must introduce at least two new bosons and two new fermions. 
In Fig.~\ref{fig:diags-Loop}, we show box diagrams which generate the neutron portal operator. 
Note that the diagram with $u$ quark in the upper right will vanish due to the same reason mentioned above (see Eq.~\eqref{eq:CCandSU(3)} and discussion around that). 
Hereafter, we focus on the left diagram in Fig.~\ref{fig:diags-Loop} for simplicity. 

\begin{table}[!ht]
\begin{center}
\begin{tabular}{c|ccc|cccc}
 & ~~$u$~~ & ~~$d$~~ & ~~$\chi$~~ & ~~~$\Psi$~~~ & ~~~$\psi$~~~ & ~~~$\sigma$~~~ & ~~~$\varphi$~~~ \\ \hline
$SU(3)_C$ & $\mathbf{3}$ & $\mathbf{3}$ & $\mathbf{1}$ & ${\bf n}_{\Psi}$ & ${\bf n}_{\psi}$ & ${\bf n}_{\sigma}$ & ${\bf n}_{\psi}$ \\
$U(1)_{\rm em}$ & $+ \frac{2}{3}$ & $- \frac{1}{3}$ & $0$ & $Q_{\Psi}$ & $Q_{\psi}$ & $Q_{\sigma}$ & $Q_{\psi}$ \\ \hline
$G_{\rm dark}$ & $\mathbf{1}$ & $\mathbf{1}$ & $\mathbf{1}$ & $\mathcal{Q}_D$ & $\mathcal{Q}_D$ & $\mathcal{Q}_D$ & $\mathcal{Q}_D$ 
\end{tabular}
\caption{Charge assignments for the box diagram. 
$\Psi$ and $\psi$ are dark fermions, and $\sigma$ and $\varphi$ are dark scalars. 
$Q_{\psi}, Q_{\sigma}$ depend on the position of the $u$ quark in the diagram, and we show the details in the main text. 
$Q_{\Psi}$ is a free parameter, and $\mathcal{Q}_D$ includes all charge assignment of $G_{\rm dark}$. 
For allowed choices of $({\bf n}_{\Psi}, {\bf n}_{\psi}, {\bf n}_{\sigma})$, see the main text. }
\label{tab:loop-box}
\end{center}
\end{table}
The charge assignments for relevant particles are shown in Table~\ref{tab:loop-box}. 
$\psi$ and $\varphi$ should have the same charges for all symmetries of the model, because they couple to $\chi$ which is singlet. 
The $U(1)_{\rm em}$ charges of $\psi, \sigma, \varphi$ depend on the position of the $u$ quark in the left diagram of Fig.~\ref{fig:diags-Loop}, which can be found as
\begin{align}
\begin{array}{c|cc}
~~~~u \text{ quark}~~~~ & ~~~\psi, \varphi~~~ & ~~~~~\sigma~~~~~ \\ \hline
\text{upper left} & Q_{\Psi} - \frac{1}{3} & Q_{\Psi} - \frac{2}{3} \\[0.3ex]
\text{lower left} & Q_{\Psi} + \frac{2}{3} & Q_{\Psi} + \frac{1}{3}
\end{array}
\label{eq:boxU1charges}
\end{align}
with given $Q_{\Psi}$. 
Moreover, $SU(3)_C$ charges of $\Psi, \psi, \sigma$ and $\varphi$ cannot be determined uniquely. 
The conditions for $({\bf n}_{\Psi}, {\bf n}_{\psi}, {\bf n}_{\sigma})$ are found as
\begin{align}
\mathbf{\bar{3}} \otimes {\bf n}_{\Psi} \otimes \overline{{\bf n}_{\sigma}} \supset \mathbf{1} \, , \quad \mathbf{\bar{3}} \otimes {\bf n}_{\sigma} \otimes \overline{{\bf n}_{\psi}} \supset \mathbf{1} \, , \quad \mathbf{\bar{3}} \otimes {\bf n}_{\psi} \otimes \overline{{\bf n}_{\Psi}} \supset \mathbf{1} \, ,
\end{align}
which are satisfied, e.g., by choosing
\begin{align}
({\bf n}_{\Psi}, {\bf n}_{\psi}, {\bf n}_{\sigma}) = (\mathbf{1}, \mathbf{3}, \mathbf{\bar{3}}) \, , ~ (\mathbf{\bar{3}}, \mathbf{1}, \mathbf{3}) \, , ~ (\mathbf{3}, \mathbf{\bar{3}}, \mathbf{1}) \, ,
\end{align}
if we restrict these to be $SU(3)_C$ singlet and (anti-)fundamental representations. 
Note that one can consider the model without $SU(3)_C$ singlet particle for $\Psi, \psi, \sigma$ and $\varphi$, e.g.,
\begin{align}
({\bf n}_{\Psi}, {\bf n}_{\psi}, {\bf n}_{\sigma}) = (\mathbf{8}, \mathbf{3}, \mathbf{6}) \, , ~ (\mathbf{6}, \mathbf{8}, \mathbf{3}) \, , ~ (\mathbf{3}, \mathbf{6}, \mathbf{8}) \, , \cdots \, .
\end{align}
It is emphasized that the $G_{\rm dark}$ charges of these four particles should be the same, because they construct the loop. 
Furthermore, thanks to these dark charges, we clearly do not have the tree-level process, and hence, the contributions from these loop diagrams become dominant. 

The relevant Lagrangian for box diagrams is
\begin{align}
- \mathcal{L} \supset \mathcal{Y}_1 \overline{q_1} \Psi \sigma^* + \mathcal{Y}_2 \overline{\Psi} q_2^c \varphi + \mathcal{Y}_3 \overline{\psi} q_3^c \sigma + \mathcal{Y}_{\chi} \overline{\chi^c} \psi \varphi^* + {\rm h.c.} \, ,
\label{eq:Lag-box}
\end{align}
where $q_{1, 2, 3}$ indicate $u$ or $d$ quarks, and $(q_1, q_2, q_3) = (u_R, d_R, d_R)$ and $(d_R, u_R, d_R)$ correspond to the diagram with $u$ quark in the upper left and lower left, respectively. 
Here, we omit $SU(3)_C$ indices, because it depends on the choice of $SU(3)_C$ representations for dark particles. 
For example, the case with $({\bf n}_{\Psi}, {\bf n}_{\psi}, {\bf n}_{\sigma}) = (\mathbf{\bar{3}}, \mathbf{1}, \mathbf{3})$ gives
\begin{align}
- \mathcal{L} \supset \mathcal{Y}_1 \epsilon^{\alpha \varepsilon \lambda} \overline{q_{1 \hspace{0.1em} \alpha}} \Psi_{\varepsilon} \sigma_{\lambda}^* + \mathcal{Y}_2 \overline{\Psi^{\beta}} q_{2 \hspace{0.1em} \beta}^c \varphi + \mathcal{Y}_3 \overline{\psi} q_{3 \hspace{0.1em} \gamma}^c \sigma^{\gamma} + \mathcal{Y}_{\chi} \overline{\chi^c} \psi \varphi^* + {\rm h.c.} \, .
\label{eq:Lag-box_ex1}
\end{align}
Note that all quarks are right-handed, while the chirality of $\chi$ is still undetermined: it can be fixed once that of $\psi$ is determined. 
Therefore, for generality, we change the last term of Eq.~\eqref{eq:Lag-box_ex1} to be
\begin{align}
\mathcal{Y}_{\chi} \overline{\chi^c} \psi \varphi^* ~ \to ~ \mathcal{Y}_{\chi}^{(L)} \overline{\chi^c_L} \psi \varphi^* + \mathcal{Y}_{\chi}^{(R)} \overline{\chi^c_R} \psi \varphi^* \, ,
\label{eq:ybkrep}
\end{align}
with appropriate $SU(3)_C$ indices for general case. 
Hereafter, we call $\mathcal{Y}_{1, 2, 3}$ and $\mathcal{Y}_{\chi}^{(L,R)}$ dark Yukawa couplings, and we choose $({\bf n}_{\Psi}, {\bf n}_{\psi}, {\bf n}_{\sigma}) = (\mathbf{\bar{3}}, \mathbf{1}, \mathbf{3})$ as an explicit example. 

Now we can evaluate the box diagram in Fig.~\ref{fig:diags-Loop}. 
The amplitude from Lagrangian in Eq.~\eqref{eq:Lag-box_ex1} with Eq.~\eqref{eq:ybkrep} is
\begin{align}
&i^4 \mathcal{Y}_1 \mathcal{Y}_2 \mathcal{Y}_3 \DarkFac \epsilon^{\alpha \varepsilon \lambda} \int \! \frac{d^4 \ell}{(2 \pi)^4} \frac{\Bigl[ \overline{\chi^c} \left( \mathcal{Y}_{\chi}^{(L)} P_R + \mathcal{Y}_{\chi}^{(R)} P_L \right) i \left( \slg{p}_{\psi} + m_{\psi} \right) q_{3 \hspace{0.1em} \gamma}^c \Bigr] \Bigl[ \overline{q_{1 \hspace{0.1em} \alpha}} i \delta^{\beta}_{\varepsilon} \left( \slg{p}_{\Psi} + m_{\Psi} \right) q_{2 \hspace{0.1em} \beta}^c \Bigr] i^2 \delta^{\gamma}_{\lambda}}{(p_{\Psi}^2 - m_{\Psi}^2) (p_{\psi}^2 - m_{\psi}^2) (p_{\sigma}^2 - m_{\sigma}^2) (p_{\varphi}^2 - m_{\varphi}^2)} \nonumber \\[0.5ex]
&= \mathcal{Y}_1 \mathcal{Y}_2 \mathcal{Y}_3 \DarkFac \epsilon^{\alpha \beta \gamma} \int \! \frac{d^4 \ell}{(2 \pi)^4} \frac{m_{\Psi} \Bigl[ \overline{\chi^c} \left( \mathcal{Y}_{\chi}^{(L)} m_{\psi} + \mathcal{Y}_{\chi}^{(R)} \slg{p}_{\psi} \right) q_{3 \hspace{0.1em} \alpha}^c \Bigr] \Bigl[ \overline{q_{1 \hspace{0.1em} \beta}} q_{2 \hspace{0.1em} \gamma}^c \Bigr]}{(p_{\Psi}^2 - m_{\Psi}^2) (p_{\psi}^2 - m_{\psi}^2) (p_{\sigma}^2 - m_{\sigma}^2) (p_{\varphi}^2 - m_{\varphi}^2)} \, ,
\end{align}
where $p_x$ and $m_x$ are four momenta and masses for inner dark particle of $x = \Psi, \psi, \sigma, \varphi$, and we implicitly use the fact that $q_{1, 2, 3}$ are the right-handed fields. 
Here, $\DarkFac$ is a numerical factor originated from the dark particle loop: for example, $\DarkFac = N_D$ when $\mathcal{Q}_D$ is a fundamental representation of $G_{\rm dark} = SU(N_D)$. 
The momentum directions for external fermions are defined that $p_{1, 2, 3}$ for $q_{1, 2, 3}$, respectively, are outgoing, while $p_4$ for $\chi$ is incoming, which leads to momentum conservation of $p_1 + p_2 + p_3 = p_4$. 
This results in the choice of momenta for propagators as, for example,
\begin{align}
p_{\Psi} = \ell - p_1 \, , \quad p_{\psi} = \ell + p_3 \, , \quad p_{\sigma} = \ell \, , \quad p_{\varphi} = \ell - p_1 - p_2 \, .
\end{align}
After the usual loop calculation which is summarized in appendix~\ref{app:boxcalc}, we get
\begin{align}
&i \frac{\mathcal{Y}_1 \mathcal{Y}_2 \mathcal{Y}_3 m_{\Psi}}{16 \pi^2 M_D^4} \DarkFac \epsilon^{\alpha \beta \gamma} \biggl[ \biggr. \mathcal{Y}_{\chi}^{(L)} m_{\psi} \left( \overline{\chi^c} q_{3 \hspace{0.1em} \alpha}^c \right) \left( \overline{q_{1 \hspace{0.1em} \beta}} q_{2 \hspace{0.1em} \gamma}^c \right) I_1 + \mathcal{Y}_{\chi}^{(R)} \left( \overline{\chi^c} \slg{p}_3 q_{3 \hspace{0.1em} \alpha}^c \right) \left( \overline{q_{1 \hspace{0.1em} \beta}} q_{2 \hspace{0.1em} \gamma}^c \right) I_2 \nonumber \\[0.3ex]
&\hspace{9.6em} + \mathcal{Y}_{\chi}^{(R)} \left( \overline{\chi^c} \slg{p}_2 q_{3 \hspace{0.1em} \alpha}^c \right) \left( \overline{q_{1 \hspace{0.1em} \beta}} q_{2 \hspace{0.1em} \gamma}^c \right) I_3 + \mathcal{Y}_{\chi}^{(R)} \left( \overline{\chi^c} \slg{p}_4 q_{3 \hspace{0.1em} \alpha}^c \right) \left( \overline{q_{1 \hspace{0.1em} \beta}} q_{2 \hspace{0.1em} \gamma}^c \right) I_4 \biggl. \biggr] \, , \label{eq:result-box}
\end{align}
where $M_D$ is a typical mass scale of dark particles, which is called a dark scale hereafter, and $I_{1, 2, 3, 4}$ are the loop functions defined as
\begin{align}
&I_1 \equiv \int \! d x_{(4)} \frac{1}{\Delta_4^2} \, , \quad I_2 \equiv \int \! d x_{(4)} \frac{x_0}{\Delta_4^2} \, , \quad I_3 \equiv \int \! d x_{(4)} \frac{- x_1}{\Delta_4^2} \, , \quad I_4 \equiv \int \! d x_{(4)} \frac{x_1 + x_3}{\Delta_4^2} \, , \\[0.5ex]
&\hspace{2.5em} \text{with } ~ \int \! d x_{(4)} \equiv \int_0^1 \! d x_0 d x_1 d x_2 d x_3 \delta \left( 1 - x_0 - x_1 - x_2 - x_3 \right) \, , \label{eq:Ix4} \\[0.3ex]
&\hspace{5.2em} \Delta_4 \equiv x_0 \frac{m_{\sigma}^2}{M_D^2} + x_1 \frac{m_{\Psi}^2}{M_D^2} + x_2 \frac{m_{\psi}^2}{M_D^2} + x_3 \frac{m_{\varphi}^2}{M_D^2} - x_0 x_3 \frac{(p_1 + p_2)^2}{M_D^2} - x_1 x_2 \frac{(p_1 + p_3)^2}{M_D^2} \nonumber \\[0.3ex]
&\hspace{8.0em} - x_0 x_1 \frac{p_1^2}{M_D^2} - x_1 x_3 \frac{p_2^2}{M_D^2} - x_0 x_2 \frac{p_3^2}{M_D^2} - x_2 x_3 \frac{p_4^2}{M_D^2} \, . \label{eq:Delta4def}
\end{align}
$\slg{p}_{2, 3, 4}$ can be replaced by corresponding external fermion masses by the Dirac equation, and when $m_{\psi}, m_{\chi} \gg m_{u, d}$, the dominant contribution in Eq.~\eqref{eq:result-box} becomes
\begin{align}
i \frac{\mathcal{Y}_1 \mathcal{Y}_2 \mathcal{Y}_3 m_{\Psi}}{16 \pi^2 M_D^4} \DarkFac \Bigl[ \mathcal{Y}_{\chi}^{(L)} m_{\psi} I_1 + \mathcal{Y}_{\chi}^{(R)} m_{\chi} I_4 \Bigr] \times \epsilon^{\alpha \beta \gamma} \left( \overline{\chi^c} q_{3 \hspace{0.1em} \alpha}^c \right) \left( \overline{q_{1 \hspace{0.1em} \beta}} q_{2 \hspace{0.1em} \gamma}^c \right) \, .
\end{align}
As a result, we obtain
\begin{align}
\left. \frac{1}{\Lambda_n^2} \right|_{\rm loop} \simeq \frac{\mathcal{Y}_1 \mathcal{Y}_2 \mathcal{Y}_3 m_{\Psi}}{16 \pi^2 M_D^4} \DarkFac \Bigl[ \mathcal{Y}_{\chi}^{(L)} m_{\psi} I_1 + \mathcal{Y}_{\chi}^{(R)} m_{\chi} I_4 \Bigr] \, .
\label{eq:Lamn-loop}
\end{align}
If we assume $m_{\Psi}^2 = m_{\psi}^2 = m_{\sigma}^2 = m_{\varphi}^2 = M_D^2 \gg m_{\chi}^2, m_{u, d}^2, (p_1 + p_{2, 3})^2$,\footnote{$(p_1 + p_2)^2$ and $(p_1 + p_3)^2$ can be considered as the Mandelstam variables, $s = (p_1 + p_2)^2 = (p_3 - p_4)^2$ and $t = (p_1 + p_3)^2 = (p_2 - p_4)^2$. 
Roughly speaking, $s, t$ are at most $\mathcal{O} (m_{\chi}^2)$ from the fact that $s + t + u = m_u^2 + 2 m_d^2 + m_{\chi}^2$, and therefore, this assumption can be applied if $M_D$ is much larger than $m_{\chi}$.} the loop functions are estimated as
\begin{align}
I_1 \sim \int \! d x_{(4)} \frac{1}{(x_0 + x_1 + x_2 + x_3)^2} = \frac{1}{6} \, , \quad I_4 \sim \int \! d x_{(4)} \frac{x_1 + x_3}{(x_0 + x_1 + x_2 + x_3)^2} = \frac{1}{12} \, ,
\end{align}
and hence, we have
\begin{align}
\left. \frac{1}{\Lambda_n^2} \right|_{\rm loop} &\sim \frac{\mathcal{Y}_1 \mathcal{Y}_2 \mathcal{Y}_3}{96 \pi^2 M_D^2} \DarkFac \left[ \mathcal{Y}_{\chi}^{(L)} + \mathcal{Y}_{\chi}^{(R)} \frac{m_{\chi}}{2 M_D} \right] \nonumber \\[0.5ex]
&\simeq \frac{1}{(17.8 \, {\rm TeV})^2} \left( \frac{\DarkFac}{3} \right) \left( \frac{1 \, {\rm TeV}}{M_D} \right)^2 \left( \frac{\mathcal{Y}_1 \mathcal{Y}_2 \mathcal{Y}_3 \mathcal{Y}_{\chi}^{(L)}}{1} \right) \nonumber \\[0.3ex]
&\hspace{1.2em} + \frac{1}{(795 \, {\rm TeV})^2} \left( \frac{\DarkFac}{3} \right) \left( \frac{1 \, {\rm TeV}}{M_D} \right)^3 \left( \frac{m_{\chi}}{1 \, {\rm GeV}} \right) \left( \frac{\mathcal{Y}_1 \mathcal{Y}_2 \mathcal{Y}_3 \mathcal{Y}_{\chi}^{(R)}}{1} \right) \, . \label{eq:Lamn-loop1}
\end{align}
If the dark particles have some mass hierarchy, we can obtain the result for $1 / \Lambda_n^2$ by omitting the corresponding term in Eq.~\eqref{eq:Delta4def}: for example, when $m_{\psi}^2 \ll m_{\Psi}^2, m_{\sigma}^2, m_{\varphi}^2 \approx M_D^2$, we can estimate each loop integral by $\Delta_4 \sim x_0 + x_1 + x_3$. 
We emphasize that the above result for the dominant contribution to $\Lambda_n$ from the box diagram in Eqs.~\eqref{eq:Lamn-loop} and \eqref{eq:Lamn-loop1} can be used for the other cases, $({\bf n}_{\Psi}, {\bf n}_{\psi}, {\bf n}_{\sigma}) = (\mathbf{1}, \mathbf{3}, \mathbf{\bar{3}}) \, , ~ (\mathbf{3}, \mathbf{\bar{3}}, \mathbf{1})$. 

From the dominant part of Eq.~\eqref{eq:Lamn-loop1}, desired $\Lambda_n$ which is $2 \, {\rm TeV} \lesssim \Lambda_n \lesssim 15 \, {\rm TeV}$ can be obtained by setting $113 \, {\rm GeV} \lesssim M_D \lesssim 845 \, {\rm GeV}$ with $\DarkFac = 3$ and $\mathcal{O} (1)$ dark Yukawa couplings. 
Therefore, larger dark Yukawa couplings are required for heavier $M_D$.\footnote{When $\Psi$ and/or $\psi$ have the same $SU(3)_C \times U(1)_{\rm em}$ charges as the SM up- and/or down-type quarks, their masses are constrained to be heavier than $1.5 \text{--} 3.0 \, {\rm TeV}$ by the 4th generation quark searches at the LHC~\cite{CMS:2020ttz,CMS:2021mku,ATLAS:2022hnn,CMS:2022fck,ATLAS:2024gyc,CMS:2024xbc,ATLAS:2024zlo}. 
Although these constraints significantly depend on the specific model setup, we choose $M_D = 3 \, {\rm TeV}$ in numerical analyses, as a conservative value.\label{foot:4thgen}} 
In fact, when each dark Yukawa coupling is set to be $\mathcal{Y}_{1, 2, 3}, \mathcal{Y}_{\chi}^{(L, R)} \approx 2.97 \, (1.09)$, we have $\Lambda_n \simeq 2 \, {\rm TeV} \, (15 \, {\rm TeV})$ for $M_D \simeq 1 \, {\rm TeV}$ with $\DarkFac = 3$. 
Note that the above calculation assumes that we have one set of $(\Psi, \psi, \sigma, \varphi)$. 
If there are several numbers of flavors for dark particles, the final result of $1 / \Lambda_n^2$ can be obtained by summing up all contributions. 
As we will see later, these numbers of flavors help to reduce the scale of $\Lambda_n$.

\subsection{Connection between UV completion and dark QCD}
\label{sec:darkQCD}

In Eqs.~\eqref{eq:Lamn-tree} and \eqref{eq:Lamn-loop}, we have obtained the $\Lambda_n$ from UV completed models. 
Now we move to discuss how they are related with the information about the DS. 
For simplicity, we restrict the symmetry in the DS is a single gauge group of $SU(N_D)$, and we call this ``dark QCD" hereafter. 
Then, we will discuss the confinement scale of the dark QCD in explicit examples. 

We assume that the DM arises as a composite state of a dark QCD with $N_D$ number of colors and $\mathcal{N}_F$ number of dark quark flavors, where they are chosen such that in the deep UV, the dark QCD is governed by an IR fixed point. 
Once a part of dark particles have TeV-scale masses (whose origin will be discussed later), they are integrated out and the low-energy effective theory has a fewer number of flavors. 
The introduction of this TeV scale then causes a perturbation to the conformal field theory (CFT), and the dark QCD becomes asymptotically free. 
As the coupling strength of the dark QCD at the fixed point is not so small, the confinement scale emerges, which can naturally lie at the GeV scale, and therefore, the DM forms at the time of confinement and inherits the GeV-scale mass. 

To be more specific, we should check the beta functions of the corresponding gauge couplings. 
For this purpose, we consider all dark particles have fundamental representation of $SU(N_D)$. 
Then, by neglecting contributions from the electroweak gauge coupling and SM/dark Yukawa couplings, the beta functions for the $SU(3)_C$ and $SU(N_D)$ gauge couplings at the two-loop level can be written as
\begin{align}
\beta_s (g_s, g_d) &\equiv \frac{d g_s}{d \ln \mu} = \frac{g_s^3}{16 \pi^2} \beta_0^{(s)} + \frac{g_s^5}{(16 \pi^2)^2} \beta_1^{(ss)} + \frac{g_s^3 g_d^2}{(16 \pi^2)^2} \beta_1^{(sd)} \, , \label{eq:betas} \\[1ex]
\beta_d (g_s, g_d) &\equiv \frac{d g_d}{d \ln \mu} = \frac{g_d^3} {16 \pi^2} \beta_0^{(d)} + \frac{g_d^5}{(16 \pi^2)^2} \beta_1^{(dd)} + \frac{g_d^3 g_s^2}{(16 \pi^2)^2} \beta_1^{(ds)} \, , \label{eq:betad}
\end{align}
where $g_s$ and $g_d$ are gauge couplings of $SU(3)_C$ and $SU(N_D)$, respectively. 
\begin{table}[!t]
\centering
\begin{tabular}{c|ccc|cc}
 & ~~SM quarks~ & ~~~~$\widetilde{\psi}_D$~~~~ & ~~~~$\psi_D$~~~~ & ~~~~$\widetilde{\phi}_D$~~~~ & ~~~~$\phi_D$~~~~ \\ \hline
~$SU(3)_C$~ & $\mathbf{3}$ & $R_F$ & $\mathbf{1}$ & $R_S$ & $\mathbf{1}$ \\
$SU(N_D)$ & $\mathbf{1}$ & $R_F$ & $R_F$ & $R_S$ & $R_S$ \\ \hline
\# flavor & $N_f$ & $\widetilde{\mathcal{N}}_F$ & $\mathcal{N}_F$ & $\widetilde{\mathcal{N}}_S$ & $\mathcal{N}_S$
\end{tabular}
\caption{Particle contents relevant for the discussion about the IR fixed point. 
Here, $R_{F, S}$ correspond to representations for dark fermions and scalars, respectively, and for our purpose, we restrict them to be fundamental representations for both $SU(3)_C$ and $SU(N_D)$. }
\label{tab:chargesgen}
\end{table}
The one-loop and two-loop contributions in the model with particle contents in Table~\ref{tab:chargesgen} can be found as~\cite{Bai:2013xga,Newstead:2014jva,Ritter:2022opo,Ritter:2024sqv,Jones:1981we}
\begin{align}
\beta_0^{(s)} &= \beta_0^{\rm SM} + \frac{4}{3} T_s (R_F) N_D \widetilde{\mathcal{N}}_F + \frac{1}{3} T_s (R_S) N_D \widetilde{\mathcal{N}}_S \, , \label{eq:beta0s} \\[1ex]
\beta_1^{(ss)} &= \beta_1^{\rm SM} + \left( \frac{20}{3} C_2 (G_s) + 4 C_2^{(s)} (R_F) \right) T_s (R_F) N_D \widetilde{\mathcal{N}}_F \nonumber \\
&\hspace{4.6em} + \left( \frac{2}{3} C_2 (G_s) + 4 C_2^{(s)} (R_S) \right) T_s (R_S) N_D \widetilde{\mathcal{N}}_S \, , \label{eq:beta1ss} \\[1ex]
\beta_1^{(sd)} &= 4 C_2^{(d)} (R_F) T_s (R_F) N_D \widetilde{\mathcal{N}}_F + 4 C_2^{(d)} (R_S) T_s (R_S) N_D \widetilde{\mathcal{N}}_S \, , \label{eq:beta1sd} \\[1ex]
\beta_0^{(d)} &= \frac{4}{3} T_d (R_F) \left( \mathcal{N}_F + N_C \widetilde{\mathcal{N}}_F \right) + \frac{1}{3} T_d (R_S) \left( \mathcal{N}_S + N_C \widetilde{\mathcal{N}}_S \right) - \frac{11}{3} C_2 (G_d) \, , \label{eq:beta0d} \\[1ex]
\beta_1^{(dd)} &= \left( \frac{20}{3} C_2 (G_d) + 4 C_2^{(d)} (R_F) \right) T_d (R_F) \left( \mathcal{N}_F + N_C \widetilde{\mathcal{N}}_F \right) \nonumber \\
&\hspace{1.2em} + \left( \frac{2}{3} C_2 (G_d) + 4 C_2^{(d)} (R_S) \right) T_d (R_S) \left( \mathcal{N}_S + N_C \widetilde{\mathcal{N}}_S \right) - \frac{34}{3} C_2^2 (G_d) \, , \label{eq:beta1dd} \\[1ex]
\beta_1^{(ds)} &= 4 C_2^{(s)} (R_F) T_d (R_F) N_C \widetilde{\mathcal{N}}_F + 4 C_2^{(s)} (R_S) T_d (R_S) N_C \widetilde{\mathcal{N}}_S \, , \label{eq:beta1ds}
\end{align}
with $N_C = 3$ being the color factor of $SU(3)_C$. 
Note that if the model has additional $\hat{\mathcal{N}}_F$ flavors of fermions and $\hat{\mathcal{N}}_S$ flavors of scalars whose charges are $SU(3)_C$ fundamental and $SU(N_D)$ singlet which are totally irrelevant to the neutron portal operator, $\beta_0^{(s)}$ and $\beta_1^{(ss)}$ are modified by $N_D \widetilde{\mathcal{N}}_{F, S} \to \hat{\mathcal{N}}_{F, S} + N_D \widetilde{\mathcal{N}}_{F, S}$. 
The other constants from the $SU(3)_C$ and $SU(N_D)$ gauge groups are
\begin{align}
T_{s, d} (R) = \frac{1}{2} \, , \quad C_2^{(s, d)} (R) = \frac{N_{C, D}^2 - 1}{2 N_{C, D}} \, , \quad C_2 (G_{s, d}) = N_{C, D} \, ,
\end{align}
for a fundamental fermion ($R = R_F$) and scalar ($R = R_S$). 
The SM beta functions for the QCD gauge coupling are
\begin{align}
\beta_0^{\rm SM} &= \frac{4}{3} T_s (R_F) N_f - \frac{11}{3} C_2 (G_s) = \frac{2}{3} N_f - 11 \, , \label{eq:beta1sSMNf} \\[1ex]
\beta_1^{\rm SM} &= \left( \frac{20}{3} C_2 (G_s) + 4 C_2^{(s)} (R_F) \right) T_s (R_F) N_f - \frac{34}{3} C_2^2 (G_s) = \frac{38}{3} N_f - 102 \, , \label{eq:beta2ssSMNf}
\end{align}
for $N_f$ quark flavors. 

For a concrete calculation of the IR fixed point and running of the gauge couplings $g_{s, d}$, we need to determine details of the model. 
In order to find a connection with $\Lambda_n$, we consider the model with $\Psi, \psi, \sigma$ and $\varphi$ in the DS, as introduced in section~\ref{sec:UV-1loop}. 
Each dark fermion/scalar has an individual number of flavors, denoted as $n_{\Psi, \psi, \sigma, \varphi}$ for $\Psi, \psi, \sigma, \varphi$, respectively. 
For $SU(3)_C$ charge assignments, we choose $({\bf n}_{\Psi}, {\bf n}_{\psi}, {\bf n}_{\sigma}) = (\mathbf{\bar{3}}, \mathbf{1}, \mathbf{3})$, which leads to $\widetilde{\mathcal{N}}_F = n_{\Psi}$, $\mathcal{N}_F = n_{\psi}$, $\widetilde{\mathcal{N}}_S = n_{\sigma}$ and $\mathcal{N}_S = n_{\varphi}$ in Eqs.~\eqref{eq:beta0s}-\eqref{eq:beta1ds}. 
This choice is attractive for our discussion, because $\psi$ can be lighter than the dark QCD confinement scale, when we set $Q_{\psi} = 0$. 
As discussed in Ref.~\cite{Bai:2013xga,Ritter:2022opo,Ritter:2024sqv}, this kind of light dark fermions can constitute a dark baryon, addressing the dark matter-baryon coincidence puzzle. 
Note that we can also consider the other case of $({\bf n}_{\Psi}, {\bf n}_{\psi}, {\bf n}_{\sigma}) = (\mathbf{1}, \mathbf{3}, \mathbf{\bar{3}})$ and $Q_{\Psi} = 0$ so that $\Psi$ plays a role of the light dark fermion. 
However, from Eq.~\eqref{eq:Lamn-loop}, $\Lambda_n$ in the loop-level process will be large due to $\Lambda_n \propto 1 / \sqrt{m_{\Psi}}$ and conflict with some phenomenological bounds discussed in section~\ref{Sec:phenomenology}. 
Therefore, we focus on the case with a light $\psi$ in this paper, although both cases will work for $\Lambda_n$ from the tree-level process in Eq.~\eqref{eq:Lamn-tree}. 

With our choice of $SU(3)_C$ representations for the dark particles, one can explicitly calculate all the relevant coefficients for the $\beta$ functions of $g_{s, d}$. 
Since $\psi$ and $\varphi$ can be light due to their charge assignments, we assume that some generations of $\psi, \varphi$ remain below the scale $M_D$, while the other generations of $\psi, \varphi$ and all generations of $\Psi, \sigma$ have a degenerate mass of $M_D$ for simplicity. 
To be specific, among the total numbers of flavors ${n}_{\psi}$ and ${n}_{\varphi}$, we set $\overline{n}_{\psi}$ and $\overline{n}_{\varphi}$ as numbers of light dark flavors for $\psi$ and $\varphi$, respectively. 
We then find the coefficients of $\beta$ functions in Eqs.~\eqref{eq:betas} and \eqref{eq:betad} for $\mu < M_D$ as
\begin{align}
\beta_0^{(s)} &= \beta_0^{\rm SM} \, , \quad \beta_1^{(ss)} = \beta_1^{\rm SM} \, , \quad \beta_1^{(sd)} = 0 \, , \label{eq:betagsCoefsLow} \\[1ex]
\beta_0^{(d)} &= \frac{2}{3} \overline{n}_{\psi} + \frac{1}{6} \overline{n}_{\varphi} - \frac{11}{3} C_2 (G_d) \, , \quad \beta_1^{(dd)} = \frac{13 N_D^2 - 3}{3 N_D} \overline{n}_{\psi} + \frac{4 N_D^2 - 3}{3 N_D} \overline{n}_{\varphi} - \frac{34}{3} C_2^2 (G_d) \, , \nonumber \\[0.5ex]
\beta_1^{(ds)} &= 0 \, . \label{eq:betagdCoefsLow}
\end{align}
For $\mu > M_D$, we should include all dark particles, and each coefficient is found as
\begin{align}
\beta_0^{(s)} &= -7 + \frac{N_D}{6} \left( 4 n_{\Psi} + n_{\sigma} \right) \, , \label{eq:betasour} \\[1ex]
\beta_1^{(ss)} &= -26 + \frac{N_D}{3} \left( 38 n_{\Psi} + 11 n_{\sigma} \right) \, , \label{eq:betassour} \\[1ex]
\beta_1^{(sd)} &= (N_D^2 - 1) \left( n_{\Psi} + n_{\sigma} \right) \, , \label{eq:betasdour} \\[1ex]
\beta_0^{(d)} &= \frac{1}{6} \left( 12 n_{\Psi} + 4 n_{\psi} + 3 n_{\sigma} + n_{\varphi} - 22 N_D \right) \, , \label{eq:betadour} \\[1ex]
\beta_1^{(dd)} &= \frac{13 N_D^2 - 3}{3 N_D} \left( 3 n_{\Psi} + n_{\psi} \right) + \frac{4 N_D^2 - 3}{3 N_D} \left( 3 n_{\sigma} + n_{\varphi} \right) - \frac{34}{3} N_D^2 \, , \label{eq:betaddour} \\[1ex]
\beta_1^{(ds)} &= 8 \left( n_{\Psi} + n_{\sigma} \right) \, , \label{eq:betadsour}
\end{align}
where we have used the fact that $N_f = 6$ in $\beta_{0, 1}^{\rm SM}$ for this energy range. 

Now we can calculate the renormalization group (RG) evolution for each of the gauge couplings $g_{s, d}$, once we fix $n_{\Psi, \psi, \sigma, \varphi}$, $\overline{n}_{\psi, \varphi}$ and input values of $g_{s, d}$ at some scale. 
For $g_s$, one can use the experimental value, $g_s = 1.217(5)$ at $\mu = M_Z$~\cite{ParticleDataGroup:2024cfk}, and its RG evolution for $\mu < M_D$ can be obtained by solving Eq.~\eqref{eq:betas}, independent of the numbers of dark flavors $n_{\Psi, \psi, \sigma, \varphi}$ and $\overline{n}_{\psi, \varphi}$. 
On the other hand, we do not have any information about the input value of $g_d$, and hence, we impose following two requirements. 
First, the confinement scale of $SU(N_D)$ is not so far but larger than that of our QCD, $\Lambda_{\rm dQCD} \gtrsim \Lambda_{\rm QCD}$. 
In our analysis, we define the confinement scale of the dark QCD by $\alpha_d (\Lambda_{\rm dQCD}) \equiv g_d (\Lambda_{\rm dQCD})^2 / (4 \pi) = \pi / 4$~\cite{Cornwall:1974vz,Peskin:1982mu}, and the same for $g_s$. 
As a concrete benchmark, we assume $\Lambda_{\rm dQCD} \approx 1 \, {\rm GeV}$. 
Under this assumption with specific values of $\overline{n}_{\psi, \varphi}$, one can get the value of $g_d$ at $\mu < M_D$: for example, with $\Lambda_{\rm dQCD} = 1 \, {\rm GeV}$, corresponding to $g_d \, (\mu = \Lambda_{\rm dQCD}) = \pi$, we respectively obtain $g_d \simeq 1.096, 1.030, 0.970$ at $\mu = 1, 3, 10 \, {\rm TeV}$ for $\overline{n}_{\psi} = 5$ and $\overline{n}_{\varphi} = 0$.\footnote{As we can see from each coefficient in Eq.~\eqref{eq:betagdCoefsLow}, the values of $g_d$ do not significantly deviate from the case with $(\overline{n}_{\psi}, \overline{n}_{\varphi}) = (5, 0)$, as long as $4 \overline{n}_{\psi} + \overline{n}_{\varphi}$ is fixed. 
For example, the case with $(\overline{n}_{\psi}, \overline{n}_{\varphi}) = (4, 4)$ leads to $g_d \simeq 1.098, 1.032, 0.971$ at $\mu = 1, 3, 10 \, {\rm TeV}$, respectively.\label{foot:lightflavs}} 
The second assumption is to have a fixed point for $g_d$ at some scale, which can be found by solving $\beta_d (g_s, g_d^*) = 0$ with $g_d^*$ being a fixed point value of $g_d$. 
Note that all coefficients $\beta_0^{(d)}, \beta_1^{(dd)}, \beta_1^{(ds)}$ cannot be zero simultaneously, and therefore, $\beta_d (g_s, g_d^*) = 0$ should be solved numerically. 
For simplicity, we assume $\beta_d (g_s, g_d^*) = 0$ at $\mu = M_D$, and hence, as observed above, $g_d^* \approx 1$ is required for $\Lambda_{\rm dQCD} \approx 1 \, {\rm GeV}$ with $M_D = \mathcal{O} (1) \, {\rm TeV}$. 

Although the two assumptions are enough to draw explicit curves of $g_{s, d}$ for fixed $n_{\Psi, \psi, \sigma, \varphi}$ and $\overline{n}_{\psi, \varphi}$, one also needs to check the behavior of $g_s$ at the high energy region. 
This is because the number of bi-fundamental dark particles, $n_{\Psi}$ and $n_{\sigma}$ in our setup, affects the $\beta_s (g_s, g_d)$, and too large $n_{\Psi}$ and/or $n_{\sigma}$ flip the sign of the $\beta$ function. 
As a result, a Landau pole potentially appears below the UV scale such as the grand unification scale. 
To avoid such a Landau pole, we maintain the feature of asymptotic freedom for $g_s$, namely, $\beta_0^{(s)} < 0$. 
Moreover, $g_d$ is also required not to have a Landau pole, which can be assured by $\beta_0^{(d)} < 0$.\footnote{Even if we solve $\beta_d (g_s, g_d^*) = 0$ at some scale, like at $\mu = M_D$, the dark gauge coupling $g_d$ at $\mu > M_D$ can deviate from $g_d^*$, due to the non-fixed point value of $g_s$ in our setup. 
Furthermore, this sign is also required for the fixed point: if we choose $\beta_0^{(d)} > 0$, $\beta_1^{(dd)}$ is always positive with $\beta_1^{(ds)} = 8 (n_{\Psi} + n_{\sigma}) > 0$, which leads to no solution for $g_d^*$. 
We have checked that one can also find solutions with IR fixed point for $g_s$. 
Although, in this case, the number of models is limited, the analysis can be done in a similar way.} 
Therefore, following conditions should be satisfied for our desired situation:
\begin{align}
4 n_{\Psi} + n_{\sigma} < \frac{42}{N_D} \, , \quad 12 n_{\Psi} + 4 n_{\psi} + 3 n_{\sigma} + n_{\varphi} < 22 N_D \, .
\label{eq:betaconds}
\end{align}
It is worth emphasizing that, if these conditions are satisfied and $g_d^* \approx 1$ at $\mu = M_D$, $g_d$ remains close to its approximate fixed-point value $g_d^*$ over a wide range of renormalization scales. 
In particular, the condition $\beta_0^{(s)} < 0$ ensures that $g_s$ decreases as the renormalization scale increases, thereby suppressing its contribution to the running of $g_d$. 
We numerically checked that, whenever a physical solution with $g_d^* \approx 1$ exists in a model with satisfying Eq.~\eqref{eq:betaconds}, the deviation of $g_d$ from $g_d^*$ at high scales remains below the $\mathcal{O} (1\%)$ level, independent of the specific choice of dark flavors. 

\begin{figure}[!t]
\begin{center}
\includegraphics[width=0.6\textwidth]{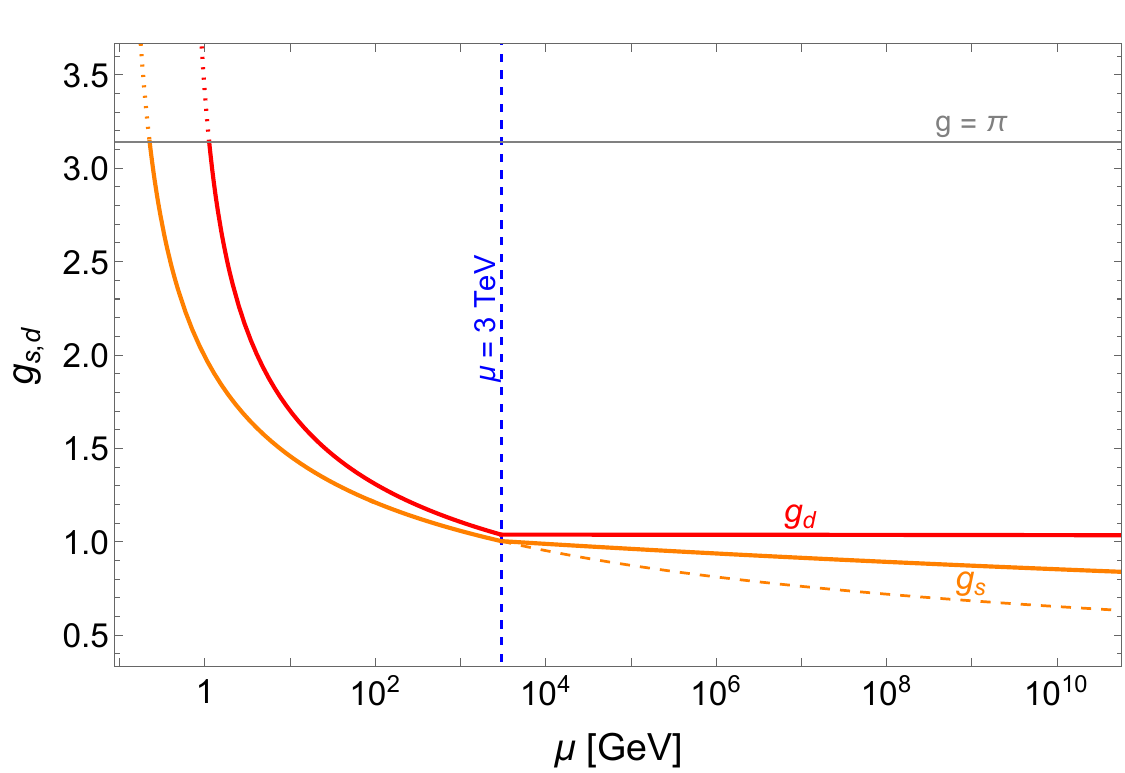}
\end{center}
\caption{The running curves of $g_s$ (orange) and $g_d$ (red), in the case of $(n_{\Psi}, n_{\psi}, n_{\sigma}, n_{\varphi}) = (2, 5, 1, 14)$ with $(\overline{n}_{\psi}, \overline{n}_{\varphi}) = (4, 4)$ as light dark flavors. 
The dotted orange line is the curve without dark particles, which is the SM $g_s$ running. 
For this figure, we assume that all heavy dark particles appear at $\mu = 3 \, {\rm TeV}$, and only $g_{s, d}$ contributions to each $\beta$ function are included. 
The confinement scale for the dark QCD is found as $\Lambda_{\rm dQCD} \approx 1.1 \, {\rm GeV}$, and $g_d^* \simeq 1.04$ is stable above $3 \, {\rm TeV}$, while $g_s$ is decreasing. }
\label{fig:grunDM1}
\end{figure}

One of solutions can be found by choosing $(n_{\Psi}, n_{\psi}, n_{\sigma}, n_{\varphi}) = (2, 5, 1, 14)$ with light dark flavors of $(\overline{n}_{\psi}, \overline{n}_{\varphi}) = (4, 4)$ for $N_D = 3$, and the corresponding RG behaviors of $g_{s, d}$ are shown in Fig.~\ref{fig:grunDM1}, up to an intermediate scale, $\mu_{\rm inter} \simeq 3.3 \times 10^{10} \, {\rm GeV}$. 
We note that this specific choice of dark flavors is one representative example to demonstrate the viability of the proposed framework, and other choices can be found by following the setup and requirements explained above. 
Here, the RG running of $g_{s, d}$ is only considered, and all other effects (the weak gauge coupling and SM/dark Yukawa couplings) are omitted. 
We obtain $g_d^* \simeq 1.038$ at $\mu = 3 \, {\rm TeV}$ and $g_d (\mu_{\rm inter}) = 1.035$ whose deviation from $g_d^*$ is about $- 0.29\%$, while $g_s (\mu_{\rm inter}) \simeq 0.843$ which is larger by $32\%$ than that within the SM (dotted orange line). 
The confinement scale of the dark QCD is $\Lambda_{\rm dQCD} \approx 1.1 \, {\rm GeV}$, and with the confinement scale of the ordinary QCD, $\Lambda_{\rm QCD} \approx 0.2 \text{--} 0.3 \, {\rm GeV}$, we obtain $\Lambda_{\rm dQCD} / \Lambda_{\rm QCD} \simeq 3.7 \text{--} 5.6$. 
Note that this ratio of confinement scales leads to the naive scaling mass of a dark baryon as~\cite{Chivukula:1989qb,Nakai:2015ptz}
\begin{align}
m_{B_D} \simeq m_p \left( \frac{N_D}{3} \right) \left( \frac{\Lambda_{\rm dQCD}}{\Lambda_{\rm QCD}} \right) \approx 4 \text{--} 6 \, {\rm GeV} \, ,
\label{eq:mBDestimate}
\end{align}
and hence, the dark matter-baryon coincidence problem can be solved. 
We will discuss some features of our (lightest) dark baryon in section~\ref{Sec:darkpion}. 

It is notable that the dark Yukawa couplings can modify the above numerical results if their individual elements are of $\mathcal{O} (1)$. 
Their contributions to Eqs.~\eqref{eq:betas} and \eqref{eq:betad} first appear at the two-loop level and can be written as
\begin{align}
\Delta \beta_{s, d}^{(\mathcal{Y})} &= - \frac{g_{s, d}^3}{(16 \pi^2)^2} \sum_{\mathcal{Y}} \mathcal{C}_{\mathcal{Y}} {\rm Tr} \bigl[ \mathcal{Y}^{\dagger} \mathcal{Y} \bigr] \, , \label{eq:betadYuk}
\end{align}
where the summation runs over all relevant dark Yukawa couplings, and $\mathcal{C}_{\mathcal{Y}}$ represents the corresponding positive coefficient. 
Here, the contributions from the SM Yukawa couplings and the weak gauge coupling remain negligible in our analysis~\cite{Bai:2013xga}. 
We have numerically verified that, if all dark Yukawa couplings are fixed to 1.0 (0.7), $g_d (\mu_{\rm inter}) = 0.949$ $(0.990)$ is obtained, which corresponds to the deviation from $g_d^*$ by $-8.5\%$ ($-4.6\%$). 
The confinement scale of the dark QCD is also modified as $\Lambda_{\rm dQCD} \approx 1.23 \, {\rm GeV}$ ($1.17 \, {\rm GeV}$).\footnote{These results are obtained by assuming that all dark Yukawa couplings remain fixed at the specified values over the entire range of renormalization scales, although they are also running parameters. 
Once the RG running of the individual dark Yukawa couplings is taken into account, both the deviation from $g_d^*$ and the resulting value of $\Lambda_{\rm dQCD}$ are expected to change slightly, but not significantly. 
Since the explicit running behavior depend on the input values and structures of dark Yukawa couplings, a detailed analysis is beyond the scope of the present work and is left for future investigation.} 
Although the deviation from $g_d^*$ at the intermediate scale is larger than that in the case without dark Yukawa contributions, $\Lambda_{\rm dQCD} \sim 1.2 \, {\rm GeV}$ is still close to the desired scale and is therefore favorable for addressing the dark matter-baryon coincidence problem. 

In the present situation where there is a mass hierarchy among dark particles, the expression of $\Lambda_n$ for the loop-level process is different from that of Eq.~\eqref{eq:Lamn-loop1}. 
The expression for $\Lambda_n$ is obtained by considering proper loop integrals and summing up all contributions from each flavor of $\Psi, \psi, \sigma, \varphi$ as
\begin{align}
\left. \frac{1}{\Lambda_n^2} \right|_{\rm loop} &\sim \frac{\mathcal{Y}_1 \mathcal{Y}_2 \mathcal{Y}_3}{16 \pi^2 M_D^2} \DarkFac \, n_{\Psi} n_{\sigma} \nonumber \\[0.3ex]
&\hspace{2.0em} \times \Biggl[ \Biggr. \frac{\mathcal{Y}_{\chi}^{(L)} m_{\psi} I_1^{(1, 1)} + \mathcal{Y}_{\chi}^{(R)} m_{\chi} I_4^{(1, 1)}}{M_D} \overline{n}_{\psi} \overline{n}_{\varphi} + \frac{\mathcal{Y}_{\chi}^{(L)} m_{\psi} I_1^{(1, 0)} + \mathcal{Y}_{\chi}^{(R)} m_{\chi} I_4^{(1, 0)}}{M_D} \overline{n}_{\psi} \delta n_{\varphi} \nonumber \\[0.3ex]
&\hspace{3.6em} + \left( \mathcal{Y}_{\chi}^{(L)} I_1^{(0, 1)} + \frac{\mathcal{Y}_{\chi}^{(R)} m_{\chi} I_4^{(0, 1)}}{M_D} \right) \delta n_{\psi} \overline{n}_{\varphi} + \left( \mathcal{Y}_{\chi}^{(L)} I_1^{(0, 0)} + \frac{\mathcal{Y}_{\chi}^{(R)} m_{\chi} I_4^{(0, 0)}}{M_D} \right) \delta n_{\psi} \delta n_{\varphi} \Biggl. \Biggr] \nonumber \\[0.5ex]
&\simeq \frac{\mathcal{Y}_1 \mathcal{Y}_2 \mathcal{Y}_3 \mathcal{Y}_{\chi}^{(L)}}{96 \pi^2 M_D^2} \DarkFac \, n_{\Psi} n_{\sigma} \delta n_{\psi} \left( n_{\varphi} + 2 \overline{n}_{\varphi} \right) \nonumber \\[0.3ex]
&\hspace{1.2em} + \frac{\mathcal{Y}_1 \mathcal{Y}_2 \mathcal{Y}_3 \mathcal{Y}_{\chi}^{(R)}}{96 \pi^2 M_D^2} \DarkFac \, n_{\Psi} n_{\sigma} \frac{m_{\chi}}{M_D} \left[ I_4^{(1, 1)} \overline{n}_{\psi} \overline{n}_{\varphi} + \frac{1}{6} \overline{n}_{\psi} \delta n_{\varphi} + \frac{1}{12} \delta n_{\psi} \left( n_{\varphi} + 3 \overline{n}_{\varphi} \right) \right] \nonumber \\[0.5ex]
&\hspace{1.2em} + \frac{\mathcal{Y}_1 \mathcal{Y}_2 \mathcal{Y}_3 \mathcal{Y}_{\chi}^{(L)}}{16 \pi^2 M_D^2} \DarkFac \, n_{\Psi} n_{\sigma} \frac{m_{\psi}}{M_D} \left( I_1^{(1, 1)} \overline{n}_{\psi} \overline{n}_{\varphi} + \frac{1}{2} \overline{n}_{\psi} \delta n_{\varphi} \right) \, , \label{eq:Lamn-loop2}
\end{align}
where $\delta n_{\psi, \varphi} \equiv n_{\psi, \varphi} - \overline{n}_{\psi, \varphi}$ indicate the numbers of heavy dark flavors, and $I_{1, 4}^{(a, b)}$ ($a, b = 0, 1$) are the corresponding loop integrals, which are summarized in appendix~\ref{app:boxcalc}. 
For $M_D \gg m_{\chi} > m_{\psi}$, the first line of Eq.~\eqref{eq:Lamn-loop2} gives the dominant contribution, and we have found $\Lambda_n \simeq 7.76 \, {\rm TeV}$ for $m_{\psi} = m_{\varphi} = 500 \, {\rm MeV}$ and $m_{\chi} = 5 \, {\rm GeV}$ with all the dark Yukawa couplings being 1.\footnote{All the dark Yukawa couplings $\mathcal{Y}_{1, 2, 3, \chi}$ have matrix form when dark particles have flavors, and all elements are generally different from each other. 
In Eq.~\eqref{eq:Lamn-loop2}, however, we simply assume all elements in each matrix to be similar.} 
We checked that even when the masses for light dark particles as well as that for $\chi$ are changed, one gets $\Lambda_n = 7.3 \text{--} 8.0 \, {\rm TeV}$ for ranges of $m_{\psi} = 0.1 \text{--} 1 \, {\rm GeV}$, $m_{\varphi} = 0.1 \text{--} 20 \, {\rm GeV}$ and $m_{\chi} = 1 \text{--} 15 \, {\rm GeV}$ with $(\overline{n}_{\psi}, \overline{n}_{\varphi}) = (4, 4)$. 
We emphasize that as mentioned in footnote~\ref{foot:lightflavs}, the similar situation can be realized by fixing $4 \overline{n}_{\psi} + \overline{n}_{\varphi}$, and $(\overline{n}_{\psi}, \overline{n}_{\varphi}) = (3, 8), (2, 12)$ with $m_{\psi} = m_{\varphi} = 500 \, {\rm MeV}$ and $m_{\chi} = 5 \, {\rm GeV}$ result in $\Lambda_{\rm dQCD} = 1.09, \, 1.05 \, {\rm GeV}$ and $\Lambda_n \simeq 4.78 , \, 3.49 \, {\rm TeV} $, respectively.\footnote{We would like to emphasize that this feature for $\Lambda_{\rm dQCD}$, together with Eq.~\eqref{eq:Lamn-loop2} for $\Lambda_n$, provides a useful guideline for constructing alternative realizations of the framework. 
In particular, viable models can be constructed with a smaller number of light dark flavors, e.g., for $(\overline{n}, \overline{n}_{\varphi}) = (1, 1)$, where the single light $\psi$ constitutes the dark baryon. 
As a concrete example with this light dark flavors, choosing $(n_{\Psi}, n_{\psi}, n_{\sigma}, n_{\varphi}) = (1, 10, 2, 4)$ yields $\Lambda_{\rm dQCD} \approx 1.17 \, {\rm GeV}$ and $\Lambda_n \simeq 5.12 \, {\rm TeV}$ for the same input masses of $m_{\psi} = 500 \, {\rm MeV}$ and $m_{\chi} = 5 \, {\rm GeV}$. 
The corresponding running curves of $g_s$ and $g_d$ are qualitatively similar to those shown in Fig.~\ref{fig:grunDM1}.} 
Note that if all generations of $\psi$ have light masses, $\overline{n}_{\psi} = n_{\psi}$ and hence $\delta n_{\psi} = 0$, the dominant contribution to $\Lambda_n$ vanishes, and the resultant $\Lambda_n$ tends to be higher than $15 \, {\rm TeV}$: for $\overline{n}_{\psi} = 5$ with $\overline{n}_{\varphi} = 0 \, (1)$ case, we obtain $\Lambda_n = 96.8 \, (48.6) \, {\rm TeV}$, although these cases give the confinement scale of $\Lambda_{\rm dQCD} \approx 1 \, {\rm GeV}$. 
Therefore, it is important to have heavy generation(s) of $\psi$ together with light generation(s) of $\varphi$ for a desired value of $\Lambda_n$.

\subsection{Origin of TeV-scale mass}
\label{sec:TeVorigin}

Some DS particles have masses of $\mathcal{O} (1 \text{--} 10) \, {\rm TeV}$ so that the resultant $\Lambda_n$ is within the desired range, although this seems like somehow an \textit{ad hoc} assumption for the model. 
This scale of the DS can be, however, explained naturally. 
For example, in the Minimal Supersymmetric Standard Model (MSSM), we also encounter the similar issue, so-called ``$\mu$ problem", which is about the scale of the bi-linear term of Higgs chiral superfields ($\mu$ term) in the superpotential. 
It has been known that this $\mu$ problem can be addressed by the Giudice-Masiero mechanism~\cite{Giudice:1988yz}: the $\mu$ term in the superpotential is forbidden at the tree-level, but it arises from the K\"ahler potential. 
Suppose that we have a chiral superfield $X$ which breaks supersymmetry (SUSY), and no $\mu$ term in the superpotential by considering, e.g., $R$ symmetry. 
In this case, we can write down the following term in the K\"ahler potential:
\begin{align}
K \supset \kappa_H \frac{X^{\dagger}}{M_{\rm Pl}} H_u H_d + {\rm h.c.} \, ,
\end{align}
where $H_{u, d}$ represent Higgs chiral superfields, $\kappa_H$ is a coefficient and $M_{\rm Pl}$ is the Planck scale for example. 
Once $F$ term of $X$ acquires a non-zero VEV, $F_X$, the \textit{effective} $\mu$ term is generated as
\begin{align}
\mu_{\rm eff} = \kappa_H \frac{F_X^{\dagger}}{M_{\rm Pl}} \, ,
\end{align}
so that one can naturally relates $\mu_{\rm eff}$ with the soft mass scale of the superpartners in the MSSM $m_{\rm soft} \sim F_X / M_{\rm Pl}$. 
Applying this mechanism to a supersymmetric extension of the present setup, the masses of dark particles can be generated in a similar manner: In the K\"ahler potential, we have
\begin{align}
K \supset \kappa_{\Phi} \frac{X^{\dagger}}{M_{\rm Pl}} \Phi_{\Psi_1} \Phi_{\Psi_2} \, ,
\end{align}
where $\Phi_{\Psi_{1, 2}}$ are the corresponding chiral superfields for dark fermions $\Psi_{1, 2}$. 
Note that the term $m_{\Psi_{12}} \Phi_{\Psi_1} \Phi_{\Psi_2}$ in the superpotential, which leads to a bare mass term of $\Psi_1 \Psi_2$, can be forbidden by, e.g., $R$ symmetry. 
Here, we assume that $\Phi_{\Psi_1}$ has the same charges with those of $\Psi$ in Table~\ref{tab:loop-box}, while $\Phi_{\Psi_2}$ has the opposite ones so that $\Phi_{\Psi_1} \Phi_{\Psi_2}$ is gauge invariant. 
This term gives the mass mixing between $\Psi_{1, 2}$, which is $\kappa_{\Phi} F_X^{\dagger} / M_{\rm Pl}$. 
Therefore, it is natural to consider the mass scale of (heavy) dark fermions to be $\mathcal{O} (1) \, {\rm TeV}$ for $F_X \sim \left( 10^{11} \, {\rm GeV} \right)^2$ with $\mathcal{O} (1)$ coefficient $\kappa_{\Phi}$. 
Note that even when $\Phi_{\Psi_2}$ including $\Psi_2$ has the opposite charges to $\Psi$, the result of $\Lambda_n$ does not change: one can consider that the couplings with $\Psi_2$ at low energy are given by Eq.~\eqref{eq:Lag-box} with $\Psi \to \Psi^c$, namely,
\begin{align}
\label{eq:oppositecharges}
- \mathcal{L} \supset \mathcal{Y}_1^{(1)} \overline{q_1} \Psi_1 \sigma^* + \mathcal{Y}_2^{(1)} \overline{\Psi_1} q_2^c \varphi + \mathcal{Y}_1^{(2)} \overline{q_1} \Psi_2^c \sigma^* + \mathcal{Y}_2^{(2)} \overline{\Psi_2^c} q_2^c \varphi + {\rm h.c.} \, .
\end{align}

Let us finally comment on gauge anomaly in the SUSY extension of the current model. 
Since our model introduces new particles charged under the SM gauge symmetries, it becomes non-trivial whether gauge anomaly is properly canceled or not. 
One solution is to introduce additional chiral superfields. 
Straightforwardly, we can include the same set of partner fields for $\Psi, \psi, \sigma, \varphi$, whose charges are opposite to the corresponding ones. 
However, more minimally, as above, we can consider, e.g., $\Psi_1$ has opposite charges of $\Psi_2$ for $n_{\Psi} = 2$, and by assigning charges to $\psi, \sigma, \varphi$ in the similar manner, gauge anomaly is properly canceled. 
In this sense, the numbers of dark flavors are required to be even, and we may need to introduce two chiral superfields to the model in section~\ref{sec:darkQCD}, in order to cancel gauge anomaly originated from $\psi$ and $\sigma$. 
However, due to the same charges of $\psi$ and $\varphi$, we can consider that they are the fermion and scalar components of the same chiral supermultiplet. 
For $\sigma$, on the other hand, we need to introduce one heavy chiral superfield whose charges are opposite to those of $\sigma$. 
As a result, only one additional chiral superfield is required to cancel gauge anomaly for the SUSY extended model. 
With SUSY breaking, the supersymmetric model can be reduced to our non-SUSY setup after integrating out all the superpartners.

\section{Dark baryon DM}
\label{Sec:darkpion}

The dark QCD has a global $U(1)_D$ symmetry understood as a ``dark baryon number".\footnote{For details about the dark baryon phenomenology as well as some common features of a dark QCD, see, e.g., Refs.~\cite{Cline:2013zca,Francis:2018xjd,Appelquist:2015yfa,Kribs:2016cew,Kamada:2020buc,Garani:2021zrr} and references therein.} 
We assume that some mechanism during the PT produces a number asymmetry in $\chi$ which is of the right order of magnitude to explain the observed baryon asymmetry. 
This can be achieved by various mechanisms, as discussed in, e.g., Refs.~\cite{Fujikura:2024jto, Girmohanta:2025wcq}. 
Then, the dark quark $\psi$ also gains chemical potential due to the Yukawa coupling, $\mathcal{Y}_{\chi} \overline{\chi^c} \psi \varphi^*$ in Eq.~\eqref{eq:Lag-box}. Therefore, once the dark QCD confines, the lightest baryon made by $\psi$ will inherit this asymmetry, and will act as our DM candidate.

To illustrate, the global $U(1)_D$ charge of $\chi$ is assigned to be $1$, such that the neutron portal operator in Eq.~\eqref{Eq:portal} violates $B$ and $D$ separately but conserves the generalized baryon number $B + D$. 
The following $U(1)_D$ charge assignments then ensure that $B + D$ is conserved in the Lagrangian of Eq.~\eqref{eq:Lag-box_ex1}: $\left(-1/3,\, -2/3,\, 0,\, -1\right)$ for $\Psi$, $\sigma$, $\varphi$, and $\psi$, respectively. 
Note that $\sigma$ cannot acquire a VEV as it is color-charged, and since $\varphi$ carries no $U(1)_D$ charge, its VEV is inconsequential for $U(1)_D$ violation. 
The number asymmetry in $\chi$ can be related to its chemical potential via
\begin{equation}
\frac{n_{\chi} - n_{\bar{\chi}}}{s} = \frac{15}{2 \pi^2 g_{*S}} \left( \frac{\mu_{\chi}}{T} \right) \, ,
\label{Eq:numberMu}
\end{equation}
where $s$ and $g_{*S}$ denote the entropy density and the effective number of degrees of freedom for entropy at temperature $T$, and $\mu_{\chi}$ is the chemical potential of $\chi$. 
For a specific model of asymmetry generation, $\mu_{\chi}$ is determined by the parameters of that model, here we treat it as an input (see, e.g., the supplemental material of Ref.~\cite{Fujikura:2024jto}). 
When $\Lambda_n \lesssim 15 \, {\rm TeV}$, the neutron portal operator remains in equilibrium at GeV temperatures, enforcing a relation between the chemical potentials of $\chi$ and the SM quarks,
\begin{equation}
\sum_{i} \mu_{\chi, i}  - \mu_{u_{R, i}} - 2 \mu_{d_{R, i}} = 0 \, ,
\label{Eq:chiSMbalance}
\end{equation}
where the sum runs over the light relativistic species at GeV temperature, and $\mu_q$ denotes the chemical potential for the SM quark $q$. 
Simultaneously, the Yukawa coupling $\mathcal{Y}_{\chi} \overline{\chi^c} \psi \varphi^*$ relates the chemical potentials of $\chi$ and $\psi$ through the decay and inverse decay processes $\chi \leftrightarrow \bar{\psi} + \varphi$,
\begin{equation}
\sum_{i} \mu_{\chi, i} - \mu_{\psi,i} = 0 \, ,
\end{equation}
where the sum is again over the light species. 
This Yukawa interaction remains in equilibrium provided $\mathcal{Y}_{\chi} \gtrsim \sqrt{T_{\rm RH} / M_{\rm Pl}} \simeq 10^{-9}$, which is easily satisfied. 
The non-zero chemical potential of $\psi$ is subsequently inherited by the dark baryon DM. 
Together with the condition of electromagnetic charge neutrality, the final SM baryon asymmetry can be related to $\mu_{\chi}$ following Refs.~\cite{Harvey:1990qw,Fujikura:2024jto}. 
In the complementary regime $15 \, {\rm TeV} \lesssim \Lambda_n \lesssim 100 \, {\rm TeV}$, where the neutron portal operator is out of equilibrium at GeV temperatures, the asymmetry is instead transferred to the SM through the decay of $\chi$ to neutron before the onset of BBN. 
 
For the DM stability, due to the fact that all the particles running in the loop of the neutron portal diagram in Fig.~\ref{fig:diags-Loop} has the same charge under dark QCD, an unbroken parity can be assigned, under which all the particles running inside the loop in Fig.~\ref{fig:diags-Loop} are odd, and SM particles are even. 
Then, for odd $N_{D}$, the lightest baryon made out of $\psi$ will be parity odd, and absolutely stable, as $\Psi$ is heavier than TeV. 
For even $N_D$, the DM stability can be arranged by a combination of parity assignment and kinematical reasons. 

The symmetric component of the DM must annihilate well before the onset of structure formation. 
The DM anti-DM pair annihilates into the dark pions. 
The velocity averaged cross-section of this annihilation can be estimated as~\cite{Garani:2021zrr}
\begin{align}
\langle \sigma_{B_D} v \rangle \sim \frac{4 \pi}{m_{B_D}^2} \gtrsim 10^{-25} \, {\rm cm^3/s} \, ,
\end{align}
where the right-hand side constraint comes from requiring that only the asymmetric component is leftover well before structure formation. 
This is easily satisfied for $\Lambda_{\rm dQCD} \simeq \mathcal{O} (1) \, {\rm GeV}$. 
However, one also needs to ensure that this energy density stored into the dark pions must be injected into the visible sector, otherwise, one would overclose the Universe or be in conflict with the extra relativistic species depending on the dark pion mass. 
Hence, we must demand the existence of a portal that can make the dark pions decay to the visible sector before the BBN. 
We now discuss how this portal operator can be generated. 

\medskip
{\noindent \textbf{Appearance of an effective ALP portal operator:}}
The UV completion of the neutron portal requires TeV-scale messenger quarks $\Psi$, which are bi-fundamental under QCD and dark QCD, and also carry hypercharge. 
Integrating out $\Psi$ therefore inevitably generates portal interactions between the dark mesons and SM gauge bosons --- an operator that arises \textit{automatically} rather than being introduced by hand. 
We now examine the phenomenological viability of this portal, and whether it can simultaneously mediate the transfer of annihilation products from the symmetric DM component into the visible sector. 

Let us consider only one generation of light dark quark, denoted as $\psi$. 
In this case, the meson spectrum does not contain any light dark pions, instead it contains a heavy $\tilde{\eta}'$, associated with the spontaneous breaking of the dark axial current $J^{\mu 5}_d$. 
This axial current is anomalous under the dark QCD, as encoded by the following equation
\begin{equation}
\partial_{\mu} J^{\mu 5}_d = - \frac{g_d^2}{32 \pi^2} \epsilon^{\alpha \beta \mu \nu} G_{d, \alpha \beta}^a G_{d, \mu \nu}^a \, ,
\label{Eq:etapanomaly}
\end{equation}
where $g_d$ is the dark QCD coupling constant, and $G_{d, \mu \nu}^a$ stands for the dark QCD field-strength tensor, while $a$ represents the dark color index.\footnote{Note that for multiple generations of light dark quark $\psi_i$, there will be light dark pions, denoted as $\tilde{\pi}$. 
In the absence of a dark photon gauge symmetry, the corresponding dark isospin triplet current is non-anomalous. 
However, $\tilde{\eta}'$ and $\tilde{\pi}$ will mix due to isospin violation effects if the dark quark masses are not equal to each other. 
Therefore, similar arguments will follow, where the final effective portal interaction will be suppressed by the corresponding mixing angle.} 
The axial current $J^{\mu5}_d$ can create $\tilde{\eta}'$ from the vacuum, parametrized by
\begin{equation}
\langle 0 | J^{\mu 5}_d (x) | \tilde{\eta}' (q) \rangle = - i q^{\mu} f_{\tilde{\eta}'} \, e^{- i q \cdot x} \, ,
\label{Eq:etapcreation}
\end{equation}
where $q^{\mu}$ is the 4-momentum of the $\tilde{\eta}'$, while $f_{\tilde{\eta}'} \sim \Lambda_{\rm dQCD}$ denotes its decay constant. 
The dark QCD generated off-shell amplitude for $\tilde{\eta}' (q_{\mu}) \to \tilde{g} (p_{\mu}, c) \tilde{g} (k_{\nu}, d)$ is
\begin{equation}
i {\cal M} (\tilde{\eta}' \to \tilde{g} \tilde{g}) = i \left( \frac{g_d^2}{4 \pi^2 f_{\tilde{\eta}'}} \right) \, \epsilon^{\nu \lambda \alpha \beta} \, p_{\alpha} k_{\beta} \, \varepsilon_{\nu}^* (p) \varepsilon_{\lambda}^* (k) \delta^{cd} \, ,
\end{equation}
where $c, d$ are dark color indices for the emitted dark gluon, denoted as $\tilde{g}$. 
Ref.~\cite{Juknevich:2009gg} analyzed the following dimension-8 effective operator generated after integrating out the heavy messenger-like quarks $\Psi$ in a box-type Feynman diagram. 
The effective Lagrangian is evaluated to be
\begin{equation}
\mathcal{L}^{(8)}_{\rm eff} \supset \frac{g_d^2}{16 \pi^2 m_{\Psi}^4} \left( \frac{g_s^2}{2} G_{\mu \nu}^a G_{\rho \sigma}^a + g_{\rm Y}^2 Q_{\Psi}^2 B_{\mu \nu} B_{\rho \sigma} \right) \times \left( \frac{1}{180} \epsilon^{\mu \nu \rho \sigma} \, \epsilon^{\alpha \beta \gamma \delta} G_{d, \alpha \beta}^c G_{d, \gamma \delta}^c \right) \, ,
\label{Eq:juk}
\end{equation}
where $g_s, g_{\rm Y}$ are the QCD, and hypercharge coupling constants, respectively, $Q_{\Psi}$ is the hypercharge of $\Psi$, which can be $2/3$, or $-1/3$, $B_{\mu \nu}$, $G_{\mu \nu}^a$ are the hypercharge and SM gluon field-strength tensors, respectively. 
Eqs.~\eqref{Eq:etapanomaly}, \eqref{Eq:etapcreation} and \eqref{Eq:juk} can be used to estimate the effective portal operator mediating the process $\tilde{\eta}' \to g g$, and $\tilde{\eta}' \to \gamma \gamma$, namely
\begin{align}
\mathcal{L}_{\rm eff}^{\tilde{\eta}'} &\supset \frac{\tilde{\eta}'}{f_{\tilde{\eta}'}} \epsilon^{\mu \nu \rho \sigma} \left[ \frac{32 \pi^2}{45} \left( \frac{\alpha_{d}}{4 \pi} \right)^2 \left( \frac{m_{\tilde{\eta}'}}{m_{\Psi}} \right)^4 \right] \bigg\{ \frac{\alpha_s}{16 \pi} G_{\mu \nu}^a G_{\rho \sigma}^a + \frac{\alpha_{\rm Y}}{8 \pi} B_{\mu \nu} B_{\rho \sigma} \bigg\} \, ,
\label{Eq:etapefflag}
\end{align}
where $\alpha_i = g_i^2 / (4 \pi)$ for $i = d, s, {\rm Y}$. 

As the primary source of $\tilde{\eta}'$ mass is from the dark QCD anomaly, one can estimate that
\begin{equation}
m_{\tilde{\eta}'}^2 \simeq \frac{\Lambda_{\rm dQCD}^4}{f_{\tilde{\eta}'}^2} \simeq \Lambda_{\rm dQCD}^2 \, .
\end{equation}
It is then clear from Eq.~\eqref{Eq:etapefflag} that $\tilde{\eta}'$ acts as an axion-like particle (ALP), endowed with the following interaction Lagrangian with the visible sector,
\begin{align}
\mathcal{L}_{\rm ALP} \supset \frac{\alpha_s}{8 \pi} \frac{\tilde{\eta}'}{f_{\rm A}} G \widetilde{G} + \frac{\alpha_{\rm Y}}{4 \pi} Q_{\Psi}^2 \frac{\tilde{\eta}'}{f_{\rm A}} B \widetilde{B} \, ,
\label{Eq:effectiveALP}
\end{align}
where $f_{\rm A}$ is the effective ALP decay constant, $\widetilde{B}, \widetilde{G}$ denote the dual field-strength tensors for hypercharge and gluon, respectively. 
Matching with Eq.~\eqref{Eq:etapefflag}, we identify the effective ALP portal, with ALP mass $\sim \Lambda_{\rm dQCD}$, and effective decay constant,
\begin{align}
f_{\rm A} \simeq \frac{45}{32 \pi^2} \Lambda_{\rm dQCD} \left( \frac{m_{\Psi}}{\Lambda_{\rm dQCD}} \right)^4 \, ,
\label{Eq:feff}
\end{align}
where we have assumed the strong coupling limit $g_d \sim 4 \pi$, and simplified $m_{\tilde{\eta}'} \simeq \Lambda_{\rm dQCD}$. 
If $\Lambda_{\rm dQCD} \sim \mathcal{O} ({\rm GeV})$, and $m_{\Psi} \sim \mathcal{O} ({\rm TeV})$, we get an effective GeV scale ALP with $f_{\rm A} \sim 10^{11} \, {\rm GeV}$, resulting into ALP lifetime $\sim 1 \, {\rm s}$. 
This order estimation for $f_{\rm A}$ naively seems to be on the boundary of BBN exclusion limit, and following Ref.~\cite{Jung:2025dyo} is even in tension with the BBN prediction. 
However, a detailed calculation is necessary to make a definitive statement, especially on the initial abundance for the dark meson. 
As we have a strong supercooled phase transition around GeV scale, the pre-existing $\tilde{\eta}'$ will be diluted significantly, and one needs to re-calculate the abundance of $\tilde{\eta}'$ and its effect on BBN. 
Here, we have just noted the interesting fact that this ALP portal coupling inevitably appears as a result of the neutron portal, which may successfully act as the portal for the DM annihilation products, though detailed calculation is necessary, both for the correct evaluation for $f_{\rm A}$, and to evaluate the BBN constraint. 

Even though the effective ALP portal may serve as the portal for DM annihilation products, it can not reheat the SM after the PT to GeV temperature by itself. 
At the same time, a detailed analysis is needed to conclusively determine the feasibility of this portal. 
Therefore, one either needs $\Lambda_n \lesssim 15 \, {\rm TeV}$ such that the neutron portal operator is in equilibrium, or we need the introduction of other portal operator for this purpose, like the Higgs portal. 
Hence, in this minimal example, where only the necessary ingredients for the neutron portal operator are introduced, it is likely that the existence of Higgs portal is necessary to maintain phenomenological viability. 
To this end, we note that it is straightforward to introduce the Higgs portal in the following way. 
Let's introduce a SM and dark gauge group singlet $S$, with $\overline{\psi} i \gamma^5 \psi S$ and $\mu_S S |H|^2$ coupling. 
Once $S$ is integrated out, one obtains the operator $(\mu_S / m_S^2) \overline{\psi} i \gamma^5 \psi |H|^2$. 
After dark QCD confinement, this operator gives rise to term like
\begin{align}
\mathcal{L}_{H \tilde{\pi}} \supset \frac{\mu_S}{m_S^2} \frac{\Lambda_{\rm dQCD}^2}{4 \pi} \tilde{\pi} |H|^2 \, ,
\end{align}
which results into a mixing angle
\begin{align}
\theta_{h \tilde{\pi}} \simeq \frac{v \mu_S}{m_S^2} \frac{\Lambda_{\rm dQCD}^2}{4 \pi} \frac{1}{m_h^2 - m_S^2} \, ,
\end{align}
where $v$, $m_h$ are the Higgs vacuum expectation value and mass, respectively. 
For $\tilde{\pi}$ mass around $500 \, {\rm MeV}$, all laboratory constrained can be satisfied if $10^{-7} \lesssim |\theta_{h \tilde{\pi}}| \lesssim 10^{-4}$, while also ensuring $\tilde{\pi}$ decays before BBN~\cite{Winkler:2018qyg}. 
This can be easily satisfied for a large range of $\mu_S$, and $m_S$, given $\Lambda_{\rm dQCD} \sim {\rm GeV}$. 
Note that dark sector parity is broken here allowing the dark pion to mix with the Higgs.

\section{Neutron portal phenomenology}
\label{Sec:phenomenology}

In this section, we delve into the phenomenological aspects of the neutron portal operator, including its effect in cosmology and prospects of laboratory searches.

\subsection{Constraints from cosmology}
The neutron portal operator, Eq.~\eqref{Eq:portal} has profound phenomenological consequences. 
Depending on $m_{\chi}$, there are two scenarios: (i) $m_{\chi} < m_n$, (ii) $m_{\chi} > m_n$. 
The former case will lead to a new neutron decay channel which was analyzed in the context of the neutron lifetime anomaly~\cite{Fornal:2018eol,McKeen:2020oyr}. 
One can also take the second case where $\chi$ could decay to SM quarks through this operator. 
The mass range $|\Delta m| \sim \mathcal{O} (100) \, {\rm MeV}$ ($\Delta m = m_{\chi} - m_n$) has been explored in Ref.~\cite{McKeen:2020oyr}. 
In this paper, we focus on the second case and extend the mass difference to $\mathcal{O} ({\rm GeV})$ as we are interested in the scenario where $\chi$ has a number asymmetry, and whose decay via the neutron portal operator transfers this to the visible baryons, explaining the observed baryon asymmetry. 

\subsubsection{$\chi$ decay modes}

We first present the possible decay channels of $\chi$ depending on $m_{\chi}$, and get the expression of lifetime of $\chi$ that is crucial for the phenomenological discussion. 
Below $\Lambda_{\rm QCD}$ temperature scale, $\chi$ can decay to hadronic states. 
We can write down the chiral perturbation formalism~\cite{Davoudiasl:2014gfa,Claudson:1981gh},
\begin{align}
\mathcal{L}_c = \delta_n \bar{\chi} \left[ n_R - \frac{i}{f_{\pi}} \left( \frac{p_R}{\sqrt{2}} \pi^- + \frac{n_R}{2} ( \sqrt{3} \eta - \pi^0) \right) \right] \, ,
\end{align}
where $\delta_n = \beta_n / \Lambda_n^2$, $\beta_n = 0.0120(26) \, {\rm GeV}^3$~\cite{Aoki:2008ku}, and $f_{\pi} = 0.0922 \, {\rm GeV}$. 
Here we include the contributions from the $\pi$ meson channels, which will be shown below to provide the dominant contributions, as well as a representative contribution from the heavier $\eta$ meson channel to illustrate the suppression of heavy meson effects. 

Below pion threshold, there are two electromagnetic channels for small $m_{\chi}$: $\chi \to n \gamma, \, p e^- \bar{\nu}$. 
The three body decay is suppressed by the phase space factor compared with two body decay. 
The two-body decay rate is given by~\cite{Davoudiasl:2014gfa}
\begin{align}
\Gamma (\chi \to n + \gamma) = \frac{\alpha \delta_n^2 F_2(0)^2}{16 m_p^2 m_{\chi}^3} (m_{\chi}^4 - m_n^4) \, ,
\end{align}
where $\alpha = 1 / 137$ is the fine-structure constant and $F_2(0) = -1.91$ is the form factor of the dipole interaction~\cite{ParticleDataGroup:2024cfk}. 
Since $\chi$ is $SU(3)_C$ singlet, there is no single gluon decay channel for $\chi$, while the 3-body decay channel involving two gluons for larger $\chi$ mass turns out to be sub-leading. 

For $m_{\chi} > m_n + m_{\pi}$, new channels $\chi \to n + \pi^0$ and $\chi \to p + \pi^-$ will become dominant. 
These decay rates are given by~\cite{Davoudiasl:2014gfa}
\begin{align}
\Gamma (\chi \to n + \pi^0) &= \frac{\delta_n^2 |\vec{p}_{\pi^0}|}{64 \pi f_{\pi}^2 m_{\chi}^2} \Bigl[ \Bigr. \left( A(m_{\chi}, m_n)^2 + B(m_{\chi}, m_n)^2 \right) f(m_{\chi}, m_n, m_{\pi^0}) \nonumber \\[0.3ex]
&\hspace{6.3em} + \left( B(m_{\chi}, m_n)^2 - A(m_{\chi}, m_n)^2 \right) m_{\chi} m_n \Bigl. \Bigr] \, , \label{Eq:nrates} \\[1.2ex]
\Gamma (\chi \to p + \pi^-) &= \frac{\delta_n^2 |\vec{p}_{\pi^-}|}{32 \pi f_{\pi}^2 m_{\chi}^2} \Bigl[ \Bigr. \left( A(m_{\chi}, m_p)^2 + B(m_{\chi}, m_p)^2 \right) f(m_{\chi}, m_p, m_{\pi^-}) \nonumber \\[0.3ex]
&\hspace{6.3em} + \left( B(m_{\chi}, m_p)^2 - A(m_{\chi}, m_p)^2 \right) m_{\chi} m_p \Bigl. \Bigr] \, , \label{Eq:prates}
\end{align}
where
\begin{align}
&A(m_1, m_2) = 1 + 1.27 \frac{m_1 + m_2}{m_1 - m_2} \, , \quad B(m_1, m_2) = 1 + 1.27 \frac{m_1 - m_2}{m_1 + m_2} \, , \nonumber \\[1ex]
&f(x,y,z) = \frac{x^2 - z^2 + y^2}{2} \, ,
\end{align}
and $|\vec{p}_{\pi^0}|, |\vec{p}_{\pi^-}|$ are the magnitudes of the 3-momentum for $\pi^0, \pi^-$ respectively. 
Due to isospin symmetry, including the effect of the pion wave-function, the decay width to proton is roughly two times that to the neutron. 
These two channels are dominant because $\pi$ mesons are the lightest hadronic states in two body decay. 
Other heavy meson like $\chi \to n + \eta$ or the muti-body decay like $\chi \to n + \pi \pi$ get a phase space factor suppression so they are smaller or at most are of the same order, but, for order estimation we can neglect them. 
Further, heavier mesons other than $\pi^{\pm}, K_L$ have too short lifetime to modify the BBN reaction chains that we consider later. 
The total decay width for $\chi$ is then found out by summing all the relevant partial decay widths,
\begin{align}
\frac{1}{\tau_{\chi}} = \Gamma^{\chi}_{t} = \sum_{i j} \theta (m_{\chi} - m_i - m_j) \Gamma (\chi \to i + j) \, ,
\label{Eq:tauchi}
\end{align}
where we use unit step function $\theta$ to represent an open channel. 

\subsubsection{BBN constraint}
\label{sec:BBN}

The predictions of the standard BBN are in remarkable agreement with the observed abundances of light elements, providing a stringent probe of new physics in the early Universe. 
The onset of BBN is typically associated with the freeze-out of the neutron-proton interconversion processes, determined by the condition,
\begin{equation}
\Gamma_{n \leftrightarrow p} (T_{\rm NF}) \simeq 3 H (T_{\rm NF}) \, .
\end{equation}
This occurs at a temperature $T_{\rm NF} \sim {\rm MeV}$, corresponding to a cosmic time $t_{\rm NF} \sim 0.73 \, {\rm s}$. 
At this stage, the neutron-to-proton ratio is given by
\begin{equation}
R^{\rm NF} \sim e^{- Q / T_{\rm NF}} \sim \frac{1}{6} \, ,
\end{equation}
where $Q \equiv m_n - m_p \simeq 1.29 \, {\rm MeV}$. 
However, the formation of light nuclei does not commence immediately due to the efficient photodissociation of deuterium by high-energy photons, a phenomenon known as the deuterium bottleneck~\cite{Fradette:2017sdd}. 
Only when the temperature drops to $T_{\rm DB} \sim 70 \, {\rm keV}$, corresponding to $t_{\rm DB} \sim 200 \, {\rm s}$, can deuterium survive and nucleosynthesis proceeds efficiently. 

During the interval between $t_{\rm NF}$ and $t_{\rm DB}$, the neutron-to-proton ratio is further reduced by neutron beta decay, reaching approximately $R^{\rm DB} \sim 1/7$. 
As nucleosynthesis proceeds, most neutrons are eventually bound into $^4 {\rm He}$, leading to a primordial helium mass fraction,
\begin{equation}
Y_p \simeq \frac{2 R^{\rm DB}}{1 + R^{\rm DB}} \sim 0.25 \, ,
\end{equation}
in excellent agreement with observations. 

The observed helium abundance $Y_p = 0.245(3)$~\cite{ParticleDataGroup:2024cfk} places a strong constraint on any new physics that modifies the neutron--proton ratio during the BBN epoch. 
Requiring $\delta Y_p \lesssim 0.01$ at $2 \sigma$ imposes stringent bounds on the properties of new particles that can decay into neutrons or protons in the time window $t \sim 1 \text{--} 200 \, {\rm s}$, such as the particle $\chi$ considered in this work. 
In the following, we will analytically estimate the modification to $Y_p$ induced by a non-zero initial abundance of $\chi$. 

\paragraph{Meson-induced strong interaction effects.}

Before discussing the modification of the neutron-to-proton ratio induced by $\chi$, we first examine the potential impact of meson-induced strong interactions on the thermal history. 
In contrast to Ref.~\cite{McKeen:2020oyr}, the $\chi$ particle considered here lies in a heavier mass range, such that its decay can produce hadrons, in particular mesons such as $\pi^0$ and $\pi^-$. 
These mesons can mediate additional $n \leftrightarrow p$ conversion channels beyond the standard weak interactions~\cite{Jung:2025dyo}, e.g., $\pi^- p \leftrightarrow \pi^0 n$. 
The corresponding reaction rate can be estimated as
\begin{equation}
\Gamma^{\rm strong}_{n \leftrightarrow p} \sim \langle \sigma v \rangle_{\rm strong} \, n_m \sim \langle \sigma v \rangle_{\rm strong} \frac{\Gamma_{\chi \to m} n_{\chi}}{\langle \sigma v \rangle_{\rm strong} n_{\rm b} + \Gamma_m} \, ,
\end{equation}
where we have used the quasi-static solution for the meson number density,
\begin{equation}
\dot{n}_m = \Gamma_{\chi \to m} n_{\chi} - \left( \langle \sigma v \rangle_{\rm strong} n_{\rm b} + \Gamma_m \right) n_m \simeq 0 \, ,
\end{equation}
and $n_m$ and $\Gamma_m$ denote the meson number density and decay width, respectively. 
The strong interaction cross section can be estimated as $\langle \sigma v \rangle_{\rm strong} \simeq 10 \, {\rm mb}$, while the baryon number density scales as $n_{\rm b} \sim 10^{-19} (1 \, {\rm s} / t)^{3 / 2} \, {\rm GeV}^3$. 
This leads to a reaction rate of order,
\begin{equation}
\Gamma^{\rm strong}_{n \leftrightarrow p} \sim 10^{-18} \text{--} 10^{-17} \, {\rm GeV} \sim 10^{7} \text{--} 10^{8} \, {\rm s}^{-1} \, .
\end{equation}
Such a rate can be significant only if the mesons are sufficiently long-lived. 
In practice, neutral pions ($\tau_{\pi^0} \sim 10^{-17} \, {\rm s}$) decay too rapidly to participate, while charged pions ($\tau_{\pi^-} \sim 10^{-8} \, {\rm s}$) can potentially contribute. 
If sufficiently efficient, the strong interaction could temporarily dominate over the weak interaction. 
However, the meson abundance is controlled by the decay of $\chi$, and decreases with time. 
As a result, the strong interaction eventually becomes subdominant. 

A significant modification of the thermal history would occur only if the transition time $t_{\rm tr}$, defined by
\begin{equation}
\Gamma^{\rm weak}_{n \leftrightarrow p} (t_{\rm tr}) \simeq \Gamma^{\rm strong}_{n \leftrightarrow p} (t_{\rm tr}) \, ,
\end{equation}
is later than the standard neutron freeze-out time. 
In that case, the freeze-out would be delayed, leading to a sizable enhancement of $\delta Y_p$. 
To avoid this effect, we require $t_{\rm tr} < t_{\rm NF}$. 
The weak interaction rate can be estimated as
\begin{align}
\Gamma^{\rm weak}_{n \leftrightarrow p} (t) &\simeq \langle \sigma v \rangle_{\rm weak} \cdot n_{e, \nu} \simeq 10^{-16} \left( \frac{t}{1 \, {\rm s}} \right)^{-1} \, {\rm mb} \cdot 10^{-10} \left( \frac{t}{1 \, {\rm s}} \right)^{-3/2} \, {\rm GeV}^3 \nonumber \\
&\simeq 10^{-25} \left( \frac{t}{1 \, {\rm s}} \right)^{-5/2} \, {\rm GeV} \, ,
\end{align}
while the strong interaction rate induced by $\chi$ decay is approximately
\begin{align}
\Gamma^{\rm strong}_{n \leftrightarrow p} (t) &\simeq \langle \sigma v \rangle_{\rm strong} \frac{{\rm Br} (\chi \to m) \Gamma_{\chi} n_{\chi} e^{- t / \tau_{\chi}}}{\langle \sigma v \rangle_{\rm strong}n_{\rm b} + \Gamma_m} \simeq 10 \, {\rm mb} \cdot \frac{\Gamma_{\chi}}{\Gamma_{m}} 10^{-19} \left( \frac{t}{1 \, {\rm s}} \right)^{-3/2} e^{- t / \tau_{\chi}} \, {\rm GeV}^3 \nonumber \\
&\simeq 10^{-26} \left( \frac{t}{1 \, {\rm s}} \right)^{-3/2} \left( \frac{\tau_{\chi}}{1 \, {\rm s}} \right)^{-1} e^{- t / \tau_{\chi}} \, {\rm GeV} \, .
\end{align}
Due to the suppression factor $(\tau_{\chi} / 1 \, {\rm s})^{-1} e^{- t / \tau_{\chi}}$, the strong interaction never dominates over the weak interaction during the BBN epoch in the parameter space of our interest. 
Therefore, meson-induced strong interactions do not impose a significant constraint on $\tau_{\chi}$ in our scenario. 
This is due to the fact that the number density of $\chi$ is governed by its asymmetry, and thus is of the same order as the SM baryons, which is much smaller than the photon number density during BBN. 

\paragraph{Modification of $Y_p$.}

Let us now estimate the modification to the primordial helium abundance $Y_p$. 
All decay channels of $\chi$ can be classified into two categories, $\Gamma_n^{\chi}$ and $\Gamma_p^{\chi}$, corresponding to final states containing neutrons or protons, respectively. 
We define the initial abundance ratio between $\chi$ and baryons as
\begin{equation}
\mathcal{A}^0 \equiv \frac{n_{\chi}^0}{n_{\rm b}^0} \, ,
\end{equation}
where the initial time refers to when $\chi$ decouples from the thermal bath and begins to decay, corresponding to $T_0 \sim \mathcal{O} (10) \, {\rm MeV}$ ($t_0 \sim \mathcal{O} (10^{-2}) \, {\rm s}$) or earlier. 
Note that in the case where $\chi$ decay generates most of the visible baryon asymmetry, $n_{\rm b}^0$ denotes any small residue amount of pre-existing baryon asymmetry that has been diluted by the PT. 
Due to the decay of $\chi$, this ratio evolves to
\begin{equation}
\mathcal{A}^{\rm NF} = \frac{n_{\chi}^{\rm NF}}{n_{\rm b}^{\rm NF}} \simeq \frac{\mathcal{A}^0}{e^{\Gamma_t^{\chi} (t_{\rm NF} - t_0)} + \mathcal{A}^0 \left( e^{\Gamma_t^{\chi} (t_{\rm NF} - t_0)} - 1 \right)} \, ,
\end{equation}
evaluated at the neutron freeze-out time $t_{\rm NF} \sim 1 \, {\rm s}$. 
Since typically $\tau_{\chi} \lesssim 1 \, {\rm s}$, most $\chi$ particles have already decayed by $t_{\rm NF}$, implying $\mathcal{A}^{\rm NF} \ll 1$. 
Therefore, only the late-time tail of $\chi$ decay can affect the neutron-to-proton ratio. 

For an analytical estimate, we approximate the remaining $\chi$ population as decaying instantaneously into neutrons and protons at $t_{\rm NF}$. 
This modifies the neutron-to-proton ratio from its standard value to
\begin{equation}
R^{\rm NF} = \frac{\Gamma_n^{\chi} / \Gamma_t^{\chi} \cdot \mathcal{A}^{\rm NF} + 1/7}{\Gamma_p^{\chi} / \Gamma_t^{\chi} \cdot \mathcal{A}^{\rm NF} + 6/7} \, .
\end{equation}
Between $t_{\rm NF}$ and the onset of nucleosynthesis at $t_{\rm DB} \sim 200 \, {\rm s}$, neutron beta decay further reduces the neutron fraction, leading to
\begin{equation}
R^{\rm DB} \simeq \frac{R^{\rm NF} e^{- \Gamma_n t_{\rm DB}}}{1 + R^{\rm NF} \left( 1 - e^{- \Gamma_n t_{\rm DB}} \right)} \, .
\label{Eq:RDB_eqn}
\end{equation}
Using the above equations, we obtain the resulting shift in the helium abundance
\begin{equation}
\delta Y_p \simeq 1.54 \times \frac{6 - \mathcal{B}}{6 (1 + \mathcal{B}) + 7 \mathcal{B} \cdot \mathcal{A}^{\rm NF}} \cdot \mathcal{A}^{\rm NF} \, ,
\label{A_dYp}
\end{equation}
where we have defined $\mathcal{B} \equiv \Gamma_p^{\chi} / \Gamma_n^{\chi}$, which depends on $m_{\chi}$ and typically lies in the range $[0, 2]$. 
It is instructive to note that the standard BBN result is recovered in the limit of $\mathcal{A}^{\rm NF} \to 0$ or $\mathcal{B} = 6$. 
The dominant suppression arises from the smallness of $\mathcal{A}^{\rm NF}$, reflecting the fact that most $\chi$ particles have already decayed before neutron freeze-out. 
For $\mathcal{A}^0 \gtrsim 1$, one finds $\Gamma_t^{\chi} \gtrsim (0.3 \, {\rm s})^{-1}$, implying that $\tau_{\chi} \lesssim \mathcal{O} (0.1) \, {\rm s}$ is a conservative and safe choice. 

The $\chi \to p + \pi^-$ decay taking place after the neutron freeze-out time can effectively modify $\mathcal{B}$ by converting protons to neutrons. 
At the same time, neutral pions decay too rapidly ($\sim 10^{-17} \, {\rm s}$) to participate in strong interactions. 
Hence, for the charged pions, the relevant processes are $p + \pi^- \to n + \pi^0$, and $\pi^- \to \mu^- + \bar{\nu}_{\mu}$, with $\tau_{\pi^-} \sim 10^{-8} \, {\rm s}$. 
Let us estimate the effective change in $\mathcal{B}$ due to this effect. 
The relevant reaction rate for the newly produced pions is
\begin{equation}
\Gamma_{\rm strong}^{\pi^-} \simeq n_p \langle \sigma v \rangle_{p \pi^- \to n \pi^0} \, ,
\end{equation}
where $\langle \sigma v \rangle \sim 10 \, {\rm mb}$ and $n_p \sim 10^{-19} (1 \, {\rm s} / t)^{3/2} \, {\rm GeV}^3$. 
This yields a conversion probability of order $\mathcal{C} = \Gamma_{\rm strong}^{\pi^-} / (\Gamma_{\rm strong}^{\pi^-} + \Gamma_{\pi^-}) \sim 3\%$, leading to a small shift in $\mathcal{B}$,
\begin{equation}
\mathcal{B} \, \to \, \frac{1 - \mathcal{C}}{1 + \mathcal{B} \mathcal{C}} \, \mathcal{B} \, .
\end{equation}
Such a correction is numerically negligible and does not affect our constraint on $\tau_{\chi}$, as discussed in appendix~\ref{app:N_BBN}. 

\paragraph{$\eta$ mismatch.}

Another important observable is the primordial deuterium abundance, $D / H$, which is strongly correlated with the baryon-to-photon ratio prior to BBN~\cite{Fields:2014uja}. 
The current measurement gives~\cite{ParticleDataGroup:2024cfk}
\begin{equation}
\left( \frac{D}{H} \right)_{\rm PDG} \simeq 2.547(29) \times 10^{-5} \, ,
\end{equation}
corresponding to a baryon-to-photon ratio
\begin{equation}
\eta_{\rm BBN} \simeq 6.040(118) \times 10^{-10} \, ,
\end{equation}
which is in excellent agreement with the value inferred from the CMB,
\begin{equation}
\eta_{\rm CMB} \simeq 6.12(4) \times 10^{-10} \, .
\end{equation}
This agreement constrains any deviation to be
\begin{equation}
\frac{\Delta \eta}{\eta} \lesssim 0.039 \, .
\end{equation}
Note that we have assumed that all the source of visible baryon asymmetry is due to $\chi$ before this point. 
Therefore, we are primarily interested in the case where $\chi$ decays prior to BBN. 
However, for the sake of a general phenomenological discussion of the neutron portal operator, we can relax this criteria. 
For example, some other DS particles can also participate in the neutron portal operator. 
We then denote $\chi$ as a representative DS state that carries an asymmetry and participates in the neutron portal. 
If $\chi$ decays after the completion of nucleosynthesis but before recombination, i.e., in the time interval $10^{3} \, {\rm s} \, \lesssim \, t \, \lesssim \, 10^{13} \, {\rm s}$, the additional baryons produced by $\chi$ decay will not be reflected in $\eta_{\rm BBN}$, but will contribute to $\eta_{\rm CMB}$. 
This leads to a mismatch between the two quantities. 
The relative shift can be estimated as
\begin{equation}
\frac{\Delta \eta}{\eta} \simeq \frac{e^{- t_{\rm NR} / \tau_{\chi}} - e^{- t_{\rm RC} / \tau_{\chi}}}{1 + 1 / \mathcal{A}^0} \, \lesssim \, 0.039 \, ,
\end{equation}
where $t_{\rm NR} \sim 10^{3} \, {\rm s}$ and $t_{\rm RC} \sim 10^{13} \, {\rm s}$. 
For $\mathcal{A}^0 \gtrsim 1$, this excludes the lifetime range,
\begin{equation}
\tau_{\chi} \in [400 \, {\rm s}, \, 10^{14} \, {\rm s}] \, .
\end{equation}
In summary, BBN considerations exclude a broad region of parameter space,
\begin{equation}
0.1 \, {\rm s} \, \lesssim \, \tau_{\chi} \, \lesssim \, 10^{13} \, {\rm s} \quad (\mathcal{A}^0 \gtrsim 1) \, .
\end{equation}
The constraint can be relaxed by reducing $\mathcal{A}^0$. 

In addition to the baryon injection, $\chi$ decay can also produce non-thermal photons, either directly or through neutral mesons. 
Although their total energy density is small compared to the thermal photon bath, high-energy photons may dissociate light nuclei. 
A detailed analysis of photodissociation effects on $Y_p$ and $D/H$ can be found in Ref.~\cite{McKeen:2020oyr}. 

\subsubsection{CMB constraint}

In the previous subsection, we derived the BBN constraint, which excludes the region $0.1 \, {\rm s} \, \lesssim \, \tau_{\chi} \, \lesssim \, 10^{13} \, {\rm s}$ for $\mathcal{A}^0 \gtrsim 1$. 
The CMB observations can further constrain longer lifetimes of $\chi$. 

If $\chi$ decays around or after recombination ($t \gtrsim 10^{13} \, {\rm s}$), the injected electromagnetically interacting particles, such as photons or $e^{\pm}$, lead to additional ionization, excitation, and heating of the intergalactic medium. 
A fraction of the injected energy is deposited into hydrogen, thereby modifying the ionization history. 
This results in an enhanced Thomson optical depth and leaves imprints on the CMB temperature and polarization anisotropies, which are tightly constrained by Planck observations. 

One possible scenario is that $\chi$ decays much later than recombination, even well beyond the age of the Universe ($\sim 10^{18} \, {\rm s}$)~\cite{McKeen:2020oyr}. 
In this case, the small fraction of $\chi$ decaying during recombination does not significantly affect the CMB. 
However, $\chi$ can no longer play the role of sharing the baryon asymmetry and instead behaves as dark matter. 
Following Refs.~\cite{Cline:2013fm,Slatyer:2012yq}, this constraint can be expressed as
\begin{equation}
\frac{f_{\rm eff} \, \epsilon_{\chi}}{\tau_{\chi}} \lesssim (10^{25} \, {\rm s})^{-1} \, ,
\end{equation}
where $\epsilon_{\chi}$ denotes the fraction of $\chi$ energy density relative to dark matter, and $f_{\rm eff}$ is the effective energy deposition efficiency, typically of order unity. 
This implies a conservative bound $\tau_{\chi} \gtrsim 10^{25} \, {\rm s}$ for a dark-matter-like $\chi$ component. 

\begin{figure}[!t]
\begin{center}
\includegraphics[width=0.48\textwidth]{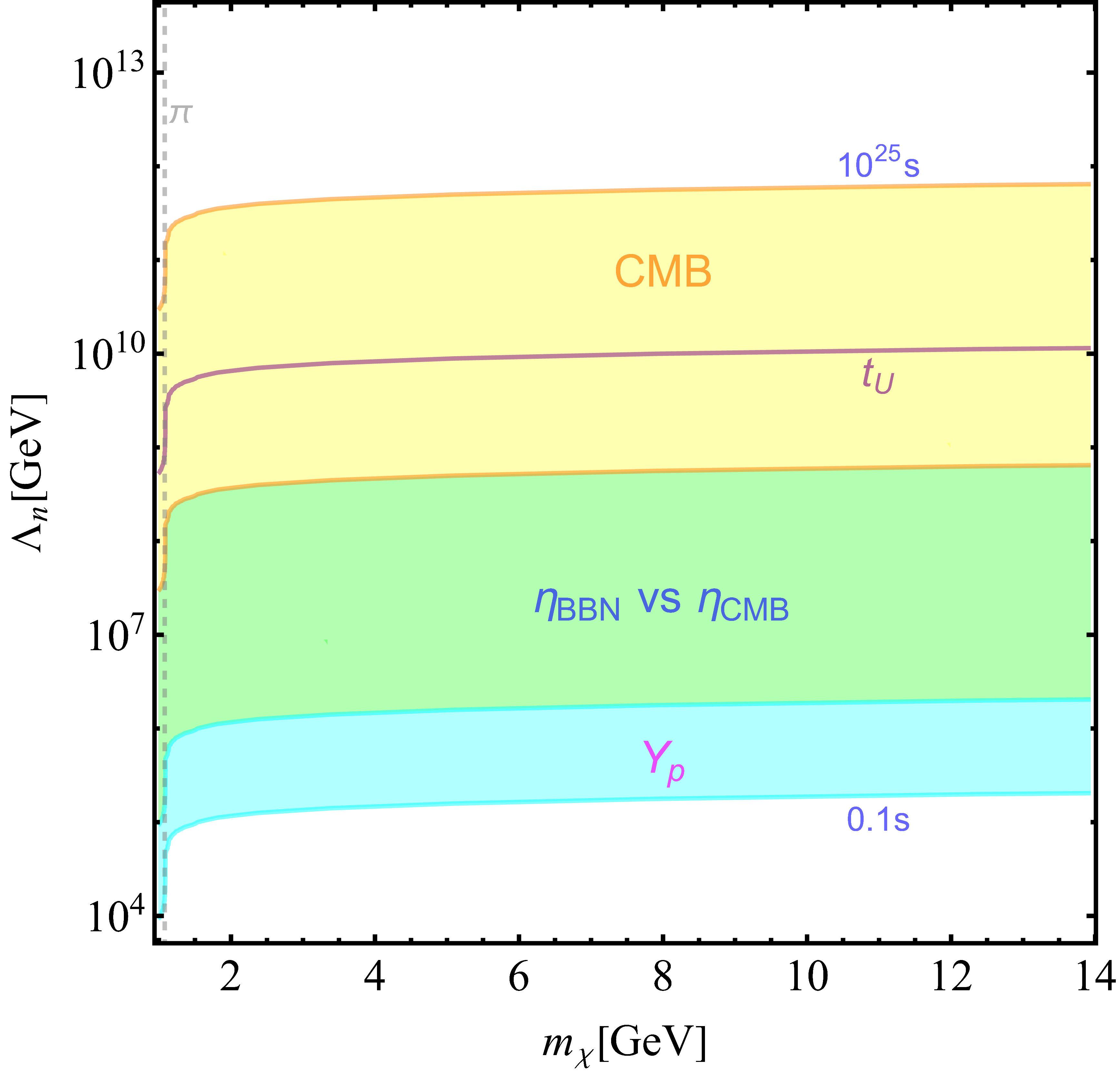}
\hspace{0.3cm}
\includegraphics[width=0.48\textwidth]{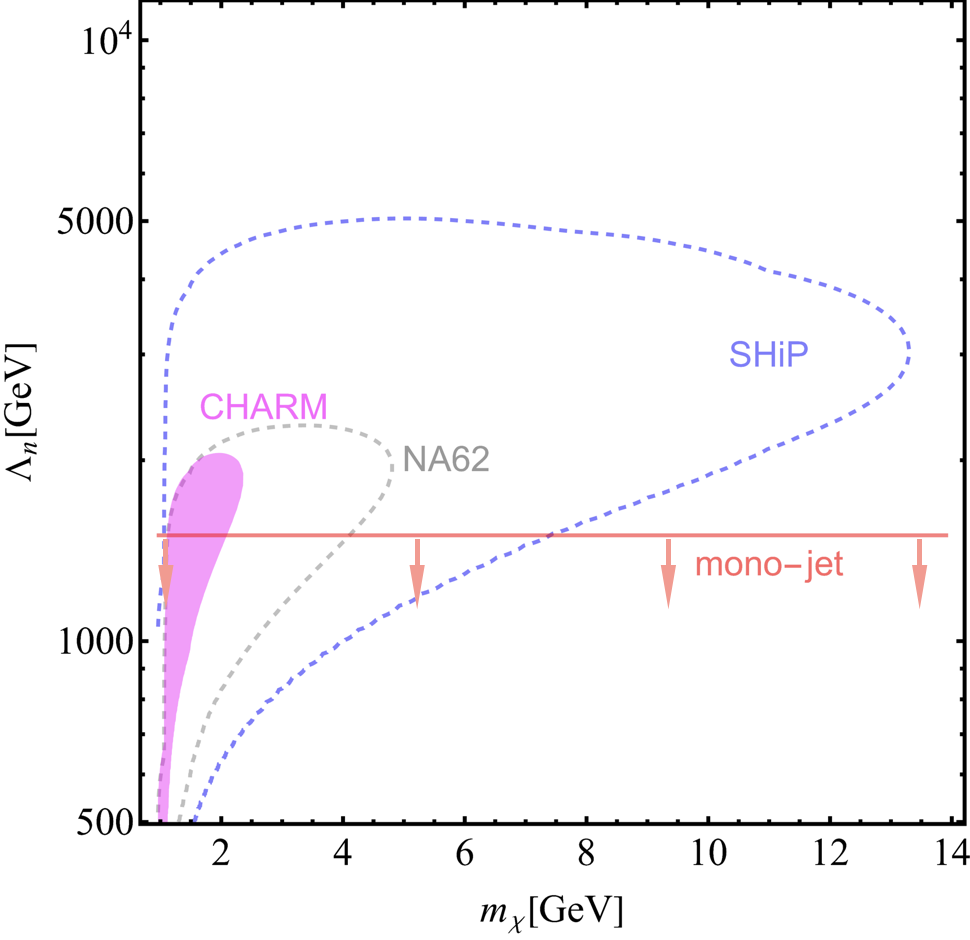}
\end{center}
\caption{\textbf{Left:} Cosmological constraints from BBN and CMB in the $(m_{\chi}, \Lambda_n)$ plane, assuming $\mathcal{A}^0 \sim 1$. 
The cyan region is excluded by the bound $\delta Y_p \lesssim 0.01$. 
The green region is excluded due to the mismatch between the baryon-to-photon ratio $\eta$ inferred from BBN and CMB, induced by $\chi$ decay. 
The yellow region is ruled out by CMB constraints. 
Long-lived $\chi$ particles are allowed only if their lifetime significantly exceeds the age of the Universe, $t_U \sim 10^{18} \, {\rm s}$. 
The vertical dashed line indicates the opening of the pion channel, which leads to a sharp change in the lifetime. 
The viable lifetime regions are $\tau_{\chi} \lesssim 0.1 \, {\rm s}$ or $\tau_{\chi} \gtrsim 10^{25} \, {\rm s}$. 
\textbf{Right:} Collider and beam dump constraints for $\Lambda_n \lesssim 10 \, {\rm TeV}$. 
The shaded region is excluded by the CHARM experiment, while the dashed contours indicate the projected sensitivities of NA62 and SHiP. 
The horizontal pink line corresponds to the mono-jet constraint, and the downward arrows indicate the excluded region below this line. 
If $\chi$ decays before BBN, the viable parameter space lies in the range $2 \, {\rm TeV} \lesssim \Lambda_n \lesssim 100 \, {\rm TeV}$. }
\label{fig:BBN}
\end{figure}

Combining the BBN and CMB constraints, the excluded region of the lifetime is approximately
\begin{equation}
0.1 \, {\rm s} \, \lesssim \, \tau_{\chi} \, \lesssim \, 10^{25} \, {\rm s} \quad (\mathcal{A}^0 \gtrsim 1) \, ,
\end{equation}
as shown in Fig~\ref{fig:BBN}. 
Therefore, for the neutron portal scenario that accounts for baryon asymmetry sharing, only the short-lifetime regime $\tau_{\chi} \lesssim 0.1 \, {\rm s}$ remains viable. 

\subsection{Beam dump and collider as probes for the neutron portal}

Existing and future proton beam-dump experiments can search for the long-lived fermion $\chi$, while the production proceeds through the neutron portal operator, which depends on $\Lambda_n$, and $m_{\chi}$ among other factors. 
Therefore, it is worthwhile to analyze the reach of current and future proton beam dump experiments in the parameter space of the neutron portal. 

To estimate the reach of beam dump experiments, one needs the information for the production cross-section of $\chi$ as the impinging proton interacts with the nucleus of the target material via the neutron portal interaction, denoted as $\sigma_{p N \to \chi + X}$, where $N$ denotes the nucleus of the target material, and $X$ stands for any other final state except $\chi$. 
After being produced, $\chi$ can decay to $n + \gamma$, or $n + \pi^0$, when the $\pi^0$ also immediately decays to $\gamma \gamma$. 
If this decay occurs within the decay volume of a given beam dump experiment, the neutron portal can be probed. 
The number of detectable events, as such, denoted as $N_{\chi}$, is estimated as follows:
\begin{align}
N_{\chi} = \frac{N_{\rm p.o.t}}{\sigma_{p N}} \int \! \frac{d \sigma_{p N \to \chi + X}}{d E_{\chi} d \theta} \, p(\ell_{\chi}) \, \varepsilon_{\rm rec} \, d E_{\chi} d \theta \, ,
\label{Eq:Nevents}
\end{align}
where $E_{\chi}$, $\theta$ denote the energy and angle of the produced $\chi$ in the laboratory frame, $N_{\rm p.o.t}$ is the number of protons on target, $\sigma_{p N}$ stands for the cross-section for the proton-nucleus scattering, $\varepsilon_{\rm rec}$ encodes factors like geometric acceptance of the detector and final state reconstruction efficiency, while $p(\ell_{\chi})$ represents the probability that the $\chi$ decays within the decay volume, and is given as
\begin{align}
p(\ell_{\chi}) = \exp \left( - \frac{d_1}{\ell_{\chi}} \right) - \exp \left( - \frac{d_2}{\ell_{\chi}} \right) \, ,
\label{Eq:pell}
\end{align}
where $d_1$, $d_2$ denote the distance from the production point to the start of the decay region and to the detector, respectively, while the decay length of $\chi$ in the laboratory frame is $\ell_{\chi} = \gamma \beta \tau_{\chi}$. 
The boost factor $\gamma = E_{\chi} / m_{\chi}$, and $\beta$ denotes the $\chi$ velocity in natural units, while the lifetime $\tau_{\chi}$ is evaluated analytically from Eq.~\eqref{Eq:tauchi}. 

To estimate the production cross-section reliably, we implemented the neutron portal operator in FeynRules~\cite{Alloul:2013bka}, and generated the parton-level cross-sections utilizing MadGraph5\_aMC~\cite{Alwall:2011uj}. 
We have implemented the neutron portal operator using the tree-level UV completion discussed in section~\ref{sec:UV-tree}, while appropriately identifying the effective scale $\Lambda_n$. 
The proton-nucleus scattering cross-section $\sigma_{pN}$ is taken from Ref.~\cite{Carvalho:2003pza,Dobrich:2015jyk} as
\begin{equation}
\sigma_{pN} = 53 \, {\rm mb} \times A^{0.77} \, ,
\end{equation}
where $A$ is the mass number of the target nucleus. 
We have considered the constraints from the CHARM experiment, the ongoing NA62 operated in beam dump mode, and the planned SHiP experiment. 
All the relevant parameters for these experiments are taken from Ref.~\cite{Winkler:2018qyg}. 
Assuming negligible background in the decay volume, the final result is shown in Fig.~\ref{fig:BBN} for $N_{\chi} \geq 3$ events, which corresponds to 95\% confidence level. 
The purple shaded region is the constraint from the NA62 experiment, while the gray and the blue dashed contours are estimated reaches for the NA62 and SHiP experiments. 
We find that the SHiP experiment can probe till $m_{\chi} \lesssim 14 \, {\rm GeV}$, while reaching $\Lambda_n \lesssim 5 \, {\rm TeV}$. 
In this analysis, we have not taken into account the detailed geometry of the detectors and their reconstruction efficiency for the relevant final states as a function of their energy and angular distribution, but represented them with the choice $\varepsilon_{\rm rec} \sim 10^{-2}$. 
This is a conservative choice, as all of these experiments employ the CERN SPS with a proton beam of $400 \, {\rm GeV}$ energy; therefore, for the relevant mass range, the produced $\chi$ will be boosted sufficiently such that its decay products lie in the forward region in the laboratory frame, where the detectors are placed. 

The neutron portal operator can induce jet + missing energy signatures via processes like $u d \to \bar{\chi} \bar{d}$, $d d \to \bar{\chi} \bar{u}$ in colliders~\cite{ATLAS:2021kxv, Ciscar-Monsalvatje:2023zkk}. 
We have checked that the numerically simulated production cross-section for $\chi$ due to $p p$ collisions at $\sqrt{s} = 13 \, {\rm TeV}$, denoted as $\sigma_{\chi}$, scales as follows
\begin{align}
\sigma_{\chi} \approx 2 \, {\rm fb} \left( \frac{10 \, {\rm TeV}}{\Lambda_n} \right)^4 \, ,
\end{align}
for $m_{\chi} = 1 \, {\rm GeV}$, and it does not depend significantly on $m_{\chi}$ for the region of interest. 
We then utilize the model-independent 95\% confidence level upper bound from the ATLAS analysis in Ref.~\cite{ATLAS:2021kxv} on the visible cross-section, defined as $\sigma_{\chi} \times {\cal A} \times \epsilon$, where ${\cal A}, \epsilon$ denote the acceptance and efficiency, respectively, to estimate a lower bound on $\Lambda_n$. 
We estimate $\Lambda_n \gtrsim 1.5 \, {\rm TeV}$ from the model-independent ATLAS bound; nevertheless, we note that a detailed analysis may improve this estimate. 
This estimated bound is shown as the pink solid line in Fig.~\ref{fig:BBN}.

\section{Conclusions}
\label{Sec:conclusion}

To solve the DM-baryon coincidence problem, an explanation for the GeV-scale of the DM is crucial in the asymmetric dark matter framework. 
On the other hand, the nano-Hz stochastic gravitational wave signal observed by the PTA can be better fitted in terms of a supercooled confining dark sector PT with reheating temperature at the GeV scale. 
Furthermore, the supercooled PT significantly dilutes any baryon asymmetry and DM abundance existing before the PT, and creating them after the PT seems attractive. 
Interestingly, a GeV-scale confining DS can naturally accommodate a self-interacting dark baryon DM, which is able to have the desired value of self-interaction cross-section, through the mediation of dark pions. 
All these coincidence requires an explanation for the emergence of the GeV-scale in a confining DS. 

We have addressed the emergence of the GeV scale through the UV completion of the neutron portal operator introduced in Eq.~\eqref{Eq:portal}, which is a necessary ingredient to reprocess the asymmetry created in the DS after the PT to the visible baryons when electroweak sphalerons have frozen out. 
The multi-TeV scale cut-off for the neutron portal operator is dynamically correlated with the emergence of the GeV-confinement scale: the DS is governed by an approximate IR fixed point, and once the DS particles introduced to UV complete the portal obtain TeV-scale masses and are integrated out, the dark QCD flows away from the fixed point and confines at the GeV scale. 

Both tree-level and loop-level UV completions have been considered. 
The tree-level completion naturally yields a low-scale cut-off amenable to collider and beam dump searches, while the loop-level completion leads to a larger cut-off that can still successfully account for baryon asymmetry sharing and satisfy all cosmological and laboratory constraints. 
Appearance of an effective ALP portal between the dark pions and SM gluons and photons are generated as a result of the new particles introduced which are charged under both SM and the DS, and we have analyzed whether the dark pions can decay to the SM before the BBN via this portal. 
Nevertheless, the need for a Higgs portal seems necessary, given the large cut-off in the loop-level UV completion and the necessity to bring the SM plasma into thermal equilibrium with the DS after the PT. 
We have outlined how to obtain such portal from renormalizable interactions. 
We have also thoroughly analyzed the phenomenological consequences of the neutron portal, including BBN and CMB constraints, and the reach of current and future beam dump experiments across the neutron portal parameter space. 

There are several avenues for future investigation. 
First, the UV completion of the neutron portal operator necessarily introduces new colored states coupled to the SM quarks and to the dark sector. 
Such states generically induce rich signatures in flavor physics. 
If the couplings contain irreducible complex phases, loop-induced electric dipole moments of the neutron, proton, electron, and nuclei can also arise, potentially providing highly sensitive probes of the portal sector. 

Second, it is highly motivated to embed the present framework into a grand unified structure. 
In our analysis, new states appearing in the UV completion were introduced in a minimal manner, but consistency with conventional grand unification suggests promoting them into complete GUT multiplets. 
Such an embedding would preserve gauge coupling unification more naturally and could reveal additional relations among portal couplings, masses, and flavor structures. 
Moreover, in our RG study of the dark gauge coupling, we focused primarily on gauge contributions and did not incorporate the effects of Yukawa couplings involving the new portal states. 
This simplification was adopted because the present work is intended as an initial study that emphasizes the core concept and demonstrates the viability of the mechanism at a qualitative level. 
We expect that including moderate Yukawa interactions would quantitatively shift the location of the IR fixed point rather than eliminate it altogether. 
Nevertheless, once messenger fields are organized into full GUT multiplets, their Yukawa interactions can modify the detailed running, thereby affecting the predicted confinement scale and the correlation between the GeV dark sector and the multi-TeV neutron portal scale. 
Extending the fixed-point analysis to include the coupled running of gauge and Yukawa interactions in a unified setup would therefore be an interesting direction for a future work. 

\vspace{0.2cm}

\textit{\textbf{Acknowledgements.---}} SG thanks Yi Chung and Tae Hyun Jung for useful discussions. 
SG acknowledges support by IBS under the project code IBS-R018-D1. 
YN is supported by Natural Science Foundation of Shanghai. 
YS is supported by Natural Science Foundation of China under grant No. W2433006.

\appendix

\section{Details of the box diagram calculation}
\label{app:boxcalc}

In this appendix, we summarize details of calculation for the box diagram in section~\ref{sec:UV-1loop}. 
We start from the original amplitude with Eqs.~\eqref{eq:Lag-box_ex1} and \eqref{eq:ybkrep}:
\begin{align}
i^4 \mathcal{Y}_1 \mathcal{Y}_2 \mathcal{Y}_3 \epsilon^{\alpha \varepsilon \lambda} \int \! \frac{d^4 \ell}{(2 \pi)^4} \frac{\Bigl[ \overline{\chi^c} \left( \mathcal{Y}_{\chi}^{(L)} P_R + \mathcal{Y}_{\chi}^{(R)} P_L \right) i \left( \slg{p}_{\psi} + m_{\psi} \right) q_{3 \hspace{0.1em} \gamma}^c \Bigr] \Bigl[ \overline{q_{1 \hspace{0.1em} \alpha}} i \delta^{\beta}_{\varepsilon} \left( \slg{p}_{\Psi} + m_{\Psi} \right) q_{2 \hspace{0.1em} \beta}^c \Bigr] i^2 \delta^{\gamma}_{\lambda}}{(p_{\Psi}^2 - m_{\Psi}^2) (p_{\psi}^2 - m_{\psi}^2) (p_{\sigma}^2 - m_{\sigma}^2) (p_{\varphi}^2 - m_{\varphi}^2)}
\end{align}
with $p_{\Psi} = \ell - p_1$, $p_{\psi} = \ell + p_3$, $p_{\sigma} = \ell$ and $p_{\varphi} = \ell - p_1 - p_2$. 
Since $q_{1, 2, 3}$ are the right-handed fields, spinor biliears in the numerator can be simplified as
\begin{align}
\overline{q_{1 \hspace{0.1em} \alpha}} \left( \slg{p}_{\Psi} + m_{\Psi} \right) q_{2 \hspace{0.1em} \beta}^c &\rightarrow \overline{q_{1 \hspace{0.1em} \alpha}} P_R \left( \slg{p}_{\Psi} + m_{\Psi} \right) P_R q_{2 \hspace{0.1em} \beta}^c = m_{\Psi} \overline{q_{1 \hspace{0.1em} \alpha}} P_R q_{2 \hspace{0.1em} \beta}^c \, , \\[0.5ex]
\overline{\chi^c} \left( \mathcal{Y}_{\chi}^{(L)} P_R + \mathcal{Y}_{\chi}^{(R)} P_L \right) \left( \slg{p}_{\psi} + m_{\psi} \right) q_{3 \hspace{0.1em} \gamma}^c &\rightarrow \overline{\chi^c} \left( \mathcal{Y}_{\chi}^{(L)} P_R + \mathcal{Y}_{\chi}^{(R)} P_L \right) \left( \slg{p}_{\psi} + m_{\psi} \right) P_R q_{3 \hspace{0.1em} \gamma}^c \nonumber \\[0.3ex]
&= \overline{\chi^c} \left( \mathcal{Y}_{\chi}^{(L)} m_{\psi} + \mathcal{Y}_{\chi}^{(R)} \slg{p}_{\psi} \right) P_R q_{3 \hspace{0.1em} \gamma}^c \, .
\end{align}
For the denominator, we can introduce the Feynman parameters as
\begin{align}
&\frac{1}{(p_{\Psi}^2 - m_{\Psi}^2) (p_{\psi}^2 - m_{\psi}^2) (p_{\sigma}^2 - m_{\sigma}^2) (p_{\varphi}^2 - m_{\varphi}^2)} \nonumber \\[0.5ex]
&\hspace{4.5em} = \int \! d x_{(4)} \frac{\Gamma (4)}{\Bigl[ x_0 (p_{\sigma}^2 - m_{\sigma}^2) + x_1 (p_{\Psi}^2 - m_{\Psi}^2) + x_2 (p_{\psi}^2 - m_{\psi}^2) + x_3 (p_{\varphi}^2 - m_{\varphi}^2) \Bigr]^4} \nonumber \\[0.3ex]
&\hspace{4.5em} = \int \! d x_{(4)} \frac{\Gamma (4)}{\left[ \left( \ell - (x_1 + x_3) p_1 - x_3 p_2 + x_2 p_3 \right)^2 - \widetilde{\Delta}_4 \right]^4} \, ,
\end{align}
where we use the definition for $\int \! d x_{(4)}$ in Eq.~\eqref{eq:Ix4}, and
\begin{align}
\widetilde{\Delta}_4 &\equiv x_0 m_{\sigma}^2 + x_1 m_{\Psi}^2 + x_2 m_{\psi}^2 + x_3 m_{\varphi}^2 - x_0 x_3 (p_1 + p_2)^2 - x_1 x_2 (p_1 + p_3)^2 \nonumber \\[0.3ex]
&\hspace{2.0em} - x_0 x_1 p_1^2 - x_1 x_3 p_2^2 - x_0 x_2 p_3^2 - x_2 x_3 p_4^2 \\[0.3ex]
&= M_D^2 \Delta_4 \, , \nonumber
\end{align}
with $\Delta_4$ defined in Eq.~\eqref{eq:Delta4def}. 
After shifting the loop momentum $\ell \to \ell' = \ell - (x_1 + x_3) p_1 - x_3 p_2 + x_2 p_3$, loop integral can be performed, and results can be found as
\begin{align}
\mathcal{Y}_1 \mathcal{Y}_2 \mathcal{Y}_3 m_{\Psi} \epsilon^{\alpha \beta \gamma} \int \! d x_{(4)} \frac{\Bigl[ \overline{\chi^c} \left( \mathcal{Y}_{\chi}^{(L)} m_{\psi} + \mathcal{Y}_{\chi}^{(R)} \{ (x_1 + x_3) \slg{p}_1 + x_3 \slg{p}_2 + (1 - x_2) \slg{p}_3 \} \right) P_R q_{3 \hspace{0.1em} \alpha}^c \Bigr] \Bigl[ \overline{q_{1 \hspace{0.1em} \beta}} P_R q_{2 \hspace{0.1em} \gamma}^c \Bigr]}{16 \pi^2 \widetilde{\Delta}_4^2} \, .
\end{align}
By using the momentum conservation and condition for Feynman parameters $x_0 + x_1 + x_2 + x_3 = 1$ under its integrals, we can simplify
\begin{align}
(x_1 + x_3) \slg{p}_1 + x_3 \slg{p}_2 + (1 - x_2) \slg{p}_3 = - x_1 \slg{p}_2 + x_0 \slg{p}_3 + (x_1 + x_3) \slg{p}_4 \, ,
\end{align}
and the result can be found as
\begin{align}
\mathcal{Y}_1 \mathcal{Y}_2 \mathcal{Y}_3 m_{\Psi} \epsilon^{\alpha \beta \gamma} \int \! d x_{(4)} &\left[ \frac{\mathcal{Y}_{\chi}^{(L)} m_{\psi} \left( \overline{\chi^c} P_R q_{3 \hspace{0.1em} \alpha}^c \right)}{\widetilde{\Delta}_4^2} + \frac{\mathcal{Y}_{\chi}^{(R)} x_0 \left( \overline{\chi^c} \slg{p}_3 P_R q_{3 \hspace{0.1em} \alpha}^c \right)}{\widetilde{\Delta}_4^2} \right. \nonumber \\[0.3ex]
&\hspace{1.0em} \left. - \frac{\mathcal{Y}_{\chi}^{(R)} x_1 \left( \overline{\chi^c} \slg{p}_2 P_R q_{3 \hspace{0.1em} \alpha}^c \right)}{\widetilde{\Delta}_4^2} + \frac{\mathcal{Y}_{\chi}^{(R)} (x_1 + x_3) \left( \overline{\chi^c} \slg{p}_4 P_R q_{3 \hspace{0.1em} \alpha}^c \right)}{\widetilde{\Delta}_4^2} \right] \left( \overline{q_{1 \hspace{0.1em} \beta}} P_R q_{2 \hspace{0.1em} \gamma}^c \right) \, . \label{eq:reskint}
\end{align}

Each loop integral can be performed numerically, and in the limit of $m_{\Psi}^2, m_{\psi}^2, m_{\sigma}^2, m_{\varphi}^2 \gg p_1^2, p_2^2, p_3^2, p_4^2, (p_1 + p_2)^2, (p_1 + p_3)^2$, we find analytical forms as follows:
\begin{align}
\int \! d x_{(4)} \frac{1}{\Delta_4^2} &\simeq \mathcal{F}_{\Psi \psi \varphi} (r_{\sigma}) + \mathcal{F}_{\psi \varphi \sigma} (r_{\Psi}) + \mathcal{F}_{\varphi \sigma \Psi} (r_{\psi}) + \mathcal{F}_{\sigma \Psi \psi} (r_{\varphi}) \, , \label{eq:I1full} \\[0.5ex]
\int \! d x_{(4)} \frac{x_1}{\Delta_4^2} &\simeq \frac{r_{\Psi}}{2 (r_{\psi} - r_{\Psi}) (r_{\varphi} - r_{\Psi}) (r_{\sigma} - r_{\Psi})} + \left( 2 + \frac{r_{\Psi}}{r_{\psi} - r_{\Psi}} + \frac{r_{\Psi}}{r_{\varphi} - r_{\Psi}} + \frac{r_{\Psi}}{r_{\sigma} - r_{\Psi}} \right) \frac{\mathcal{F}_{\psi \varphi \sigma} (r_{\Psi})}{2} \nonumber \\[1ex]
&\hspace{1.2em} + \frac{r_{\psi} \mathcal{F}_{\varphi \sigma \Psi} (r_{\psi})}{2 (r_{\psi} - r_{\Psi})} + \frac{r_{\varphi} \mathcal{F}_{\sigma \Psi \psi} (r_{\varphi})}{2 (r_{\varphi} - r_{\Psi})} + \frac{r_{\sigma} \mathcal{F}_{\Psi \psi \varphi} (r_{\sigma})}{2 (r_{\sigma} - r_{\Psi})} \, , \label{eq:Intx1}
\end{align}
with $r_x \equiv m_x^2 / M_D^2$ for $x = \Psi, \psi, \sigma, \varphi$. 
For convenience, the function $\mathcal{F}_{x y z} (r_X)$ is defined as
\begin{align}
\mathcal{F}_{x y z} (r_X) \equiv \frac{r_X \ln r_X}{(r_x - r_X) (r_y - r_X) (r_z - r_X)} \, ,
\label{eq:calFdef}
\end{align} 
where each of $x, y, z, X$ is $\Psi, \psi, \sigma$ or $\varphi$, and this is symmetric function under any exchange of $(x, y, z)$. 
Since $x_{0, 1, 2, 3}$ are dummy variables in the integrals, other integrals whose numerators are $x_0, x_2, x_3$ can be easily found by replacing $r_{\Psi}, r_{\psi}, r_{\sigma}, r_{\varphi}$ appropriately. 
For example, $\int \! d x_{(4)} x_3 / \Delta_4^2$ can be obtained by $r_{\Psi} \leftrightarrow r_{\varphi}$ in Eq.~\eqref{eq:Intx1}. 
The last term in Eq.~\eqref{eq:reskint} can be obtained by the sum of $\int \! d x_{(4)} x_1 / \Delta_4^2$ and $\int \! d x_{(4)} x_3 / \Delta_4^2$, and its analytical form is
\begin{align}
\int \! d x_{(4)} \frac{x_1 + x_3}{\Delta_4^2} &\simeq - \frac{r_{\sigma} r_{\psi} - r_{\Psi} r_{\varphi}}{2 (r_{\sigma} - r_{\Psi}) (r_{\psi} - r_{\Psi}) (r_{\sigma} - r_{\varphi}) (r_{\psi} - r_{\varphi})} \label{eq:I4full} \\[0.3ex]
&\hspace{1.2em} + \left( \frac{r_{\sigma}}{r_{\sigma} - r_{\Psi}} + \frac{r_{\sigma}}{r_{\sigma} - r_{\varphi}} \right) \frac{\mathcal{F}_{\Psi \psi \varphi} (r_{\sigma})}{2} + \left( \frac{r_{\sigma}}{r_{\sigma} - r_{\Psi}} + \frac{r_{\psi}}{r_{\psi} - r_{\Psi}} \right) \frac{\mathcal{F}_{\psi \varphi \sigma} (r_{\Psi})}{2} \nonumber \\[0.3ex]
&\hspace{1.2em} + \left( \frac{r_{\psi}}{r_{\psi} - r_{\Psi}} + \frac{r_{\psi}}{r_{\psi} - r_{\varphi}} \right) \frac{\mathcal{F}_{\varphi \sigma \Psi} (r_{\psi})}{2} + \left( \frac{r_{\sigma}}{r_{\sigma} - r_{\varphi}} + \frac{r_{\psi}}{r_{\psi} - r_{\varphi}} \right) \frac{\mathcal{F}_{\sigma \Psi \psi} (r_{\varphi})}{2} \, . \nonumber
\end{align}
In the limit of $r_{\Psi, \psi, \sigma, \varphi} \to 1$, we find
\begin{align}
\int \! d x_{(4)} \frac{1}{\Delta_4^2} &= \frac{1}{6} \, , \quad \int \! d x_{(4)} \frac{x_n}{\Delta_4^2} = \frac{1}{24} \, , \quad \int \! d x_{(4)} \frac{x_n + x_m}{\Delta_4^2} = \frac{1}{12} \, , \label{eq:intlims}
\end{align}
for $n, m = 0, 1, 2, 3$. 

For the specific case considered in section~\ref{sec:darkQCD}, we have utilized several types of loop integrals, especially in Eq.~\eqref{eq:Lamn-loop2}. 
The definitions of $I_{1, 4}^{(a, b)}$ are
\begin{align}
I_1^{(a, b)} &\equiv \int \! d x_{(4)} \frac{1}{(x_0 + x_1 + r_{\psi}^a x_2 + r_{\varphi}^b x_3)^2} \, , \label{eq:I1ab} \\[0.5ex]
I_4^{(a, b)} &\equiv \int \! d x_{(4)} \frac{x_1 + x_3}{(x_0 + x_1 + r_{\psi}^a x_2 + r_{\varphi}^b x_3)^2} \, , \label{eq:I4ab}
\end{align}
for $a, b = 0, 1$. 
Each analytical form can be obtained by taking $r_{\Psi, \sigma} \to 1$ and appropriate limit for $r_{\psi, \varphi}$ in Eqs.~\eqref{eq:I1full} and \eqref{eq:I4full}. 
It is notable that except for $(a, b) = (1, 1)$ case, we have exact (for $(a, b) = (0, 0)$ case) or approximate (for $(a, b) = (0, 1)$ and $(1, 0)$ cases) results for each integral as
\begin{align}
&I_1^{(0, 0)} = \frac{1}{6} \, , \quad \left. I_1^{(0, 1)} \right|_{r_{\varphi} \to 0} = \frac{1}{2} \, , \quad \left. I_1^{(1, 0)} \right|_{r_{\psi} \to 0} = \frac{1}{2} \, , \label{eq:I1approx} \\
&I_4^{(0, 0)} = \frac{1}{12} \, , \quad \left. I_4^{(0, 1)} \right|_{r_{\varphi} \to 0} = \frac{1}{3} \, , \quad \left. I_4^{(1, 0)} \right|_{r_{\psi} \to 0} = \frac{1}{6} \, . \label{eq:I4approx}
\end{align}
Moreover, we have simple analytical form for $(a, b) = (1, 1)$ case when $r_{\varphi} \to r_{\psi}$ as
\begin{align}
\left. I_1^{(1, 1)} \right|_{r_{\varphi} \to r_{\psi}} &= \frac{2 r_{\psi} - 2 - (1 + r_{\psi}) \ln r_{\psi}}{(1 - r_{\psi})^3} \, , \label{eq:I1degenerate} \\[0.3ex]
\left. I_4^{(1, 1)} \right|_{r_{\varphi} \to r_{\psi}} &= \frac{2 r_{\psi} - 2 - (1 + r_{\psi}) \ln r_{\psi}}{2 (1 - r_{\psi})^3} \, . \label{eq:I4degenerate}
\end{align}

\section{Numerical results related to BBN constraint}
\label{app:N_BBN}

\begin{figure}[!t]
\begin{center}
\includegraphics[width=0.6\textwidth]{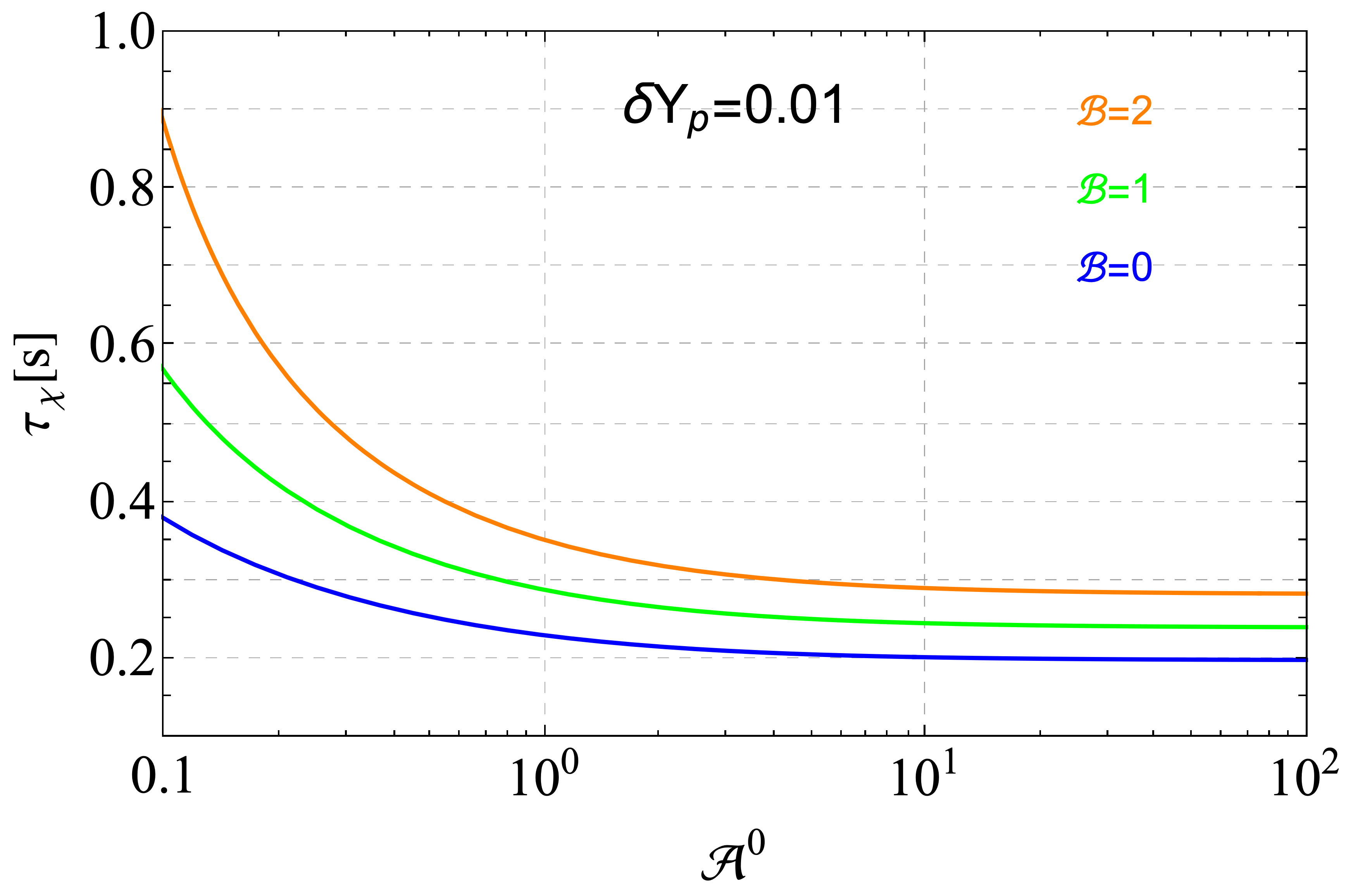}
\end{center}
\vspace{-5mm}
\caption{Relation between the initial abundance ratio $\mathcal{A}^0 = n_{\chi}^0 / n_{\rm b}^0$ and the lifetime $\tau_{\chi}$, obtained by imposing $\delta Y_p = 0.01$. 
The three curves correspond to different values of $\mathcal{B} \equiv \Gamma_p^{\chi} / \Gamma_n^{\chi} = 0, 1, 2$. 
The results are based on the analytical estimate given in Eq.~\eqref{A_dYp}. }
\label{fig:tau_chi}
\end{figure}

Here, we present numerical estimates of the BBN constraint on the lifetime $\tau_{\chi}$. 
We fix the deviation of the helium abundance to be $\delta Y_p = 0.01$, and solve for the corresponding allowed values of $\tau_{\chi}$ as a function of the initial abundance ratio $\mathcal{A}^0$. 
Fig.~\ref{fig:tau_chi} shows the resulting constraint for three representative values of $\mathcal{B} \equiv \Gamma_p^{\chi} / \Gamma_n^{\chi}$, namely $\mathcal{B} = 0, 1, 2$. 
These values effectively correspond to different regions of the $\chi$ mass parameter space. 
As can be seen from the figure, the upper bound on $\tau_{\chi}$ is only weakly sensitive to both $\mathcal{A}^0$ and $\mathcal{B}$, especially for $\mathcal{A}^0 \gtrsim \mathcal{O} (1)$. 
This indicates that the BBN constraint derived in section~\ref{sec:BBN} is robust against variations in the branching ratios and initial abundance of $\chi$.

\bibliographystyle{utphys}
\bibliography{nportal.bib}

\end{document}